\documentclass[aps,reprint,floatfix,footinbib]{revtex4-1}
\usepackage{hyperref}
\usepackage{float}
\usepackage{epsfig}
\usepackage{graphicx}
\usepackage{amsmath}
\usepackage{amsfonts}
\usepackage{amssymb}
\usepackage{bm}
\usepackage{bbold}
\usepackage{epsfig}
\usepackage{graphicx}
\usepackage{color}
\usepackage{pictex}
\usepackage{mathtools}
\usepackage{extarrows}
\usepackage{footnote}
\usepackage{footmisc}
\newcommand{\s}{\scriptscriptstyle}
\begin{document}
\title{On the phase diagram and topological order in the modulated $XYZ$ chain with magnetic fields}
\author{Toplal Pandey}
\affiliation{Department of Physics, Laurentian University, Sudbury, Ontario,
P3E 2C6 Canada}
\author{Gennady Y. Chitov}
\affiliation{Department of Physics, Laurentian University, Sudbury, Ontario,
P3E 2C6 Canada}
\date{\today}

%
%xxxxxxxxxxxxxxxxxxxxxxxxxxxxxxxxxxxxxxxxxxxxxxxxxxxxxxxxxxxxxxxxxxxxxxxxxxxxxx
%
\begin{abstract}
The $XYZ$ antiferromagnetic spin-$1/2$ chain with alternation of the exchange and anisotropy couplings  in the presence
of uniform and staggered axial magnetic fields is studied. The analysis is done using the effective quadratic fermionic Hamiltonian resulting from the Hartee-Fock approximation. Combining the exact and the mean-field methods, the local and string order parameters on the ground-state phase diagram of the model are identified and calculated. We found a topological phase with oscillating string order with a period of four lattice spacings, not reported before for this model. A detailed analysis of patterns of the string order is given.
The special $XXZ$ limit of the model with additional $U(1)$ symmetry brings about, in agreement with the Lieb-Schultz-Mattis theorem and its extensions,
plateaux of magnetization and some additional conserving quantities. We have shown that in the $XYZ$ chain, where the plateaux are smeared, the robust oscillating string order parameter is continuously  connected to its $XXZ$ limit.  Also, the non-trivial winding number and zero-energy localized Majorana edge states, as additional attributes of topological order, are robust in that phase, even off the line of $U(1)$ symmetry.
\end{abstract}
\maketitle

%
%
%%%%%%%%%%%%%%%%%%%%%%%%%%%%%%%%%%%%%%%%%%%%%%%%%%%%%%%%%%%%%%%%%%%%%%%%%%%%%%
%
\section{Introduction: Model and Context}\label{Intro}
%
%%%%%%%%%%%%%%%%%%%%%%%%%%%%%%%%%%%%%%%%%%%%%%%%%%%%%%%%%%%%%%%%%%%%%%%%%%%%%%
%
%
%
This paper is about the ground-state properties of the modulated $XYZ$ spin-$1/2$ chain.
Its Hamiltonian in the presence of uniform ($h$) and staggered ($h_a$) axial magnetic fields is:
\begin{eqnarray}
\label{XYZHam}
H &=& \sum_{n=1}^{N}\frac{J}{4}\Big[ \big(1+(-1)^n\delta\big)\big(\sigma_{n}^{x}
\sigma_{n+1}^{x}+\sigma_{n}^{y}
\sigma_{n+1}^{y}+\Delta\sigma_{n}^{z}
\sigma_{n+1}^{z}\big) \nonumber \\
&+& \big(\gamma+ (-1)^n \gamma_a \big) \big(\sigma_{n}^{x}
\sigma_{n+1}^{x}-\sigma_{n}^{y}
\sigma_{n+1}^{y}\big)\Big] \nonumber \\
&+& \frac{1}{2}\big(h+(-1)^n h_a\big)\sigma_{n}^{z}~,
\end{eqnarray}
where $\sigma$-s are the standard Pauli matrices.
The chain has bond alternation with parameter $|\delta| \leq 1$. We also allowed the $xy$ anisotropy $\gamma$ to be modulated with $\gamma_a$. In this paper we consider the antiferromagnetic ($J>0$) model at zero temperature.

The model \eqref{XYZHam} is not solvable in general, the exact solutions based on the Bethe ansatz, are available only for some special
cases. For the historical references of the isotropic $XXZ$ model with zero field, see papers by Yang and Yang \cite{*Yang:1966a,*Yang:1966-I,*Yang:1966-II,*Yang:1966-III}, for more comprehensive reviews of the available exact results see, e.g.
\cite{Takahashi:1999,McCoy:2010,Franchini:2017}, and for a most recent account of integrability and more references, see
\cite{Shiraishi:2019}.
The standard Jordan-Wigner (JW) transformation \cite{LiebSM:1961,Franchini:2017} maps \eqref{XYZHam} onto the interacting fermionic
Hamiltonian
\begin{widetext}
\begin{eqnarray}
\label{FermHam}
H &=& \sum_{n=1}^N \frac{J}{2} \big(1+(-1)^n\delta\big)
\Big[ \big(c_n^\dag c_{n+1}+\mathrm{H.c.} \big)+
2\Delta \Big(c_n^\dag c_n -\frac12 \Big) \Big(c_{n+1}^\dag c_{n+1} -\frac12 \Big) \Big]   \nonumber \\
 &+& \frac{J}{2} \big(\gamma+ (-1)^n \gamma_a \big)  \big(c_n^\dag c_{n+1}^\dag  + \mathrm{H.c.} \big)+
 \big(h+(-1)^n h_a \big) \Big(c_n^\dag c_n- \frac12 \Big)~.
\end{eqnarray}
\end{widetext}
Using spin-fermion dualities and mappings between the $XYZ$ and 8-vertex models, the isotropic $XXZ$ limit and the 6-vertex model,
den Nijs \cite{denNijs:1981} proposed the ($\gamma,\Delta$)-phase diagram of the $XYZ$ model with zero fields and modulations.
The isotropic $XXZ$ model is gapless at $|\Delta|<1$, and its perturbations by, e.g., staggered field ($h_a$), dimerization ($\delta$),
or anisotropy ($\gamma$) result in a gap opening. However, the interference of different relevant perturbations can result in their cancellations
at some values of model's parameters leading to gapless points or lines of quantum criticality. Scaling analysis of such
perturbations and their mappings onto the operators of the 8- (6-) vertex model, lead to important conclusion about  non-universality of the $XYZ$ or
$XXZ$ models \cite{Luther:1975,denNijs:1981}. The phase diagram of the $XXZ$ chain with uniform and staggered fields was proposed from
scaling analysis in Ref.~\cite{Alcaraz:1995}, see also \cite{Okamoto:1996}.
The gapless phase of the $XXZ$ model is the Luttinger liquid in fermionic language, and its transition into a gapped phase along the line of $U(1)$-symmetry $\gamma=0$ is of the Berezinskii-Kosterlitz-Thouless (BKT) class \cite{denNijs:1981,Black:1981}.

In the context of huge recent interest in topological materials and Majorana fermions \cite{Bernevig:2013,RyuSchnyder:2010,Kitaev:2001,Alicea:2012},
the fermionic Hamiltonian of type \eqref{FermHam} written more often in terms of Majorana operators, belongs to a very actively studied class of models known under the name of Kitaev-Majorana chains in recent literature. The fermionic representation \eqref{FermHam} is the chain  of interacting Majorana fermions with dimerized hopping and modulated anomalous (superconducting) pairing and chemical potential. The solvable at special symmetric points  Kitaev-Majorana  models with dimerizations and spatial modulations of potential were studied very actively in recent years with the focus on their topological phases with hidden orders and Majorana edge states \cite{Wakatsuki:2014,Ezawa:2017,Miao:2017,Chitov:2018}, similar  models in more general settings  were studied, e.g., in Refs.~\cite{Rosch:2011,Katsura:2015,Katsura:2017}. See Refs.~\cite{Alicea:2012,Wakatsuki:2014,Ezawa:2017,Miao:2017,Chitov:2018,Rosch:2011,Katsura:2015,Katsura:2017}
also for more references on quite vast literature on the models with Majorana fermions.

The non-interacting limit ($\Delta=0$) of the model \eqref{FermHam} (a.k.a. $XY$ chain) is known
to have quite rich phase diagram \cite{Perk:1975,DuttaTIM:2015,Chitov:2019}.
Very recently \cite{Chitov:2019} one the gapped phases of that model was reported to possess a hidden topological order diagnosed by
nonlocal string order parameter (SOP) \cite{denNijs:1989}, oscillating with a period of four lattice spacings.
In the view of lack of information about the modulated $XYZ$ model \eqref{XYZHam}, it is natural to explore to which extend the results of
Ref.~\cite{Chitov:2019} can be generalized for the interacting case $\Delta \neq 0$. The phase diagram of the model \eqref{XYZHam} is one of
the main results of the present study.

Another more broad goal of this work aligns with the recent effort \cite{Chitov:2017JSM,Chitov:2018,Chitov:2019} to weave  nonlocal (topological) orders into extended Landau paradigm. Technically, the key point is to incorporate string operators, string correlation functions, and SOPs \cite{denNijs:1989} into the standard framework. The local and nonlocal order parameters are related by duality, so in a sense it is a matter of choice of variables of the Hamiltonian \cite{Chitov:2017JSM,Chitov:2018,Kogut:1979,ChenHu:2007,Xiang:2007,NussinovChen:2008,*Nussinov:2013}. Another bedrock of the Landau theory is symmetry change. In the spin/fermionic systems like \eqref{XYZHam}-\eqref{FermHam} the appearance of nonlocal SOP is accompanied by the hidden $\mathbb{Z}_2 \otimes \mathbb{Z}_2$  symmetry breaking \cite{Kennedy:1992,*Kohmoto:1992,*Oshikawa:1992}.
These are internal discrete symmetries of spin reversals, and they form the Klein four-group \cite{Nomura:2015,*Nomura:2017}, a.k.a. the dihedral group
\cite{Pollmann:2012}, isomorphic to $\mathbb{Z}_2 \otimes \mathbb{Z}_2$ group.
In some cases the duality can simply map the nonlocal order onto an average of some decoupled local operator, e.g., magnetization,
and the hidden symmetry breaking becomes apparent in terms of the sublattice magnetization(s) on a dual lattice, with one or both of the Ising $\mathbb{Z}_2$ symmetries broken \cite{Xiang:2007,Chitov:2017JSM,Chitov:2018}. In general manifestations of the hidden symmetry breaking are less straightforward.

An important task addressed in this paper was to formalize the technical protocol:
In the proposed unifying formalism the role of the Ginzburg-Landau effective action is played by the effective quadratic (Hartree-Fock)
fermionic Hamiltonian. All local and nonlocal order parameters are calculated from the string correlation functions of Majorana fermions,
evaluated from the limiting values of determinants of the block Toeplitz matrices. For the quadratic Hamiltonian the elements of those
matrices are found in a closed analytical form as functions of the effective (or renormalized) couplings of the Hamiltonian. The latter are calculated from
the self-consistent minimization equations.

It appears that the notion of topological order itself is not understood uniquely in the literature. In connection to the spin chain, it appears to be associated  to the additional $U(1)$ symmetry of its isotropic $XXZ$ limit. In such limit, the Lieb-Schultz-Mattis (LSM) theorem \cite{LiebSM:1961} and its subsequent generalizations \cite{Affleck:1997,Oshikawa:2000,Nomura:2015,*Nomura:2017} predict either gapless incommensurate phase without symmetry breaking, or gapped phases with broken $\mathbb{Z}_2 \otimes \mathbb{Z}_2$ symmetry, integer fillings,  and plateaux of magnetization.  The plateaux are sometimes viewed as a hallmark of topological order. Our understanding of topological order is not tied up to the continuous $U(1)$ symmetry or related plateaux.
The gapped phases with broken (discreet) symmetry are secured by the extension of the LSM theorem for the spin chains
without continuous symmetry \cite{Tasaki:2019}. At $\gamma \neq 0$ the plateaux are smeared, but the robust SOP still exists and is continuously  connected to its $\gamma=0$ limit. Thus we associate topological order with a non-trivial SOP.
Also, the non-trivial winding number and zero-energy localized Majorana edge states, as additional attributes of topological order, are robust in the topological phase even aside from the line of $U(1)$ symmetry, in agreement with analogous exact results \cite{Chitov:2018,Chitov:2019}.

The rest of the paper is organized as follows: In Sec.~\ref{D0Res} we present a concise account of exact results for the non-interacting limit of the model: spectrum, phase diagram, and some average quantities. Those are building blocks to be used in the effective Hamiltonian and in the mean-field equations. Sec.~\ref{MFA} presents the derivation of the mean-field equations and renormalized parameters.
Sec.~\ref{Res} contains the results for the $XYZ$ chain. We present the phase diagram, local and nonlocal order parameters, winding numbers for each phase. Sec.~\ref{XXZRes} presents the results for the isotropic $XXZ$ limit of the model. Since more analytical work can be done in this limit,
more qualitative discussions of the results are presented, including the role of interaction, robustness of the mean-field approximation, and relation of the reported topological order to earlier findings of the spontaneous magnetism in this model. The algebraically ordered incommensurate gapless phase is analysed in this section as well. The results are summarized and discussed in the concluding Sec.~\ref{Concl}.
%
%
%
%xxxxxxxxxxxxxxxxxxxxxxxxxxxxxxxxxxxxxxxxxxxxxxxxxxxxxxxxxxxxxxxxxxxxxxxxxxxxxx
%
\section{Non-interacting limit $\Delta=0$}\label{D0Res}
%
%xxxxxxxxxxxxxxxxxxxxxxxxxxxxxxxxxxxxxxxxxxxxxxxxxxxxxxxxxxxxxxxxxxxxxxxxxxxxxx
%
%
%
%
%
%%%%%%%%%%%%%%%%%%%%%%%%%%%%%%%%%%%%%%%%%%%%%%%%%%%%%%%%%%%%%%%%%%%%%%%%%%%%%%
\subsection{Spectrum and phase diagram}\label{SpecPDiag}
%%%%%%%%%%%%%%%%%%%%%%%%%%%%%%%%%%%%%%%%%%%%%%%%%%%%%%%%%%%%%%%%%%%%%%%%%%%%%%
%
%
In the non-interacting limit $\Delta \equiv J_z/J =0$ the model \eqref{XYZHam} is exactly-solvable.  It was first introduced and analyzed by
Perk \textit{et al} \cite{Perk:1975}. See also \cite{Fei:2002,*Fei:2003,deLima:2007,Divakaran:2008,Chitov:2017JSM} for related
more recent work on different versions of the model. The most recent comprehensive analysis of the ground-state phase diagram of the model at $\gamma_a=0$ and its local and nonlocal order parameters is given in \cite{Chitov:2019}. It turns out that
introducing alternation of anisotropy $\gamma_a$ does not change the results \cite{Chitov:2019} qualitatively, resulting in some
minor modifications which we present below. The non-interacting results are used in the subsequent analysis of the case $\Delta \neq 0$.
We will always assume $|\gamma_a| <|\gamma|$ and from now on we set $J=1$.
We also modify for further convenience the hopping term of  the Hamiltonian \eqref{FermHam} as
\begin{equation}
\label{t0}
  1+(-1)^n\delta \longmapsto  t+(-1)^n\delta~.
\end{equation}
Referring readers to \cite{Chitov:2019} for technical details, in this section we present a concise account of the results for $\gamma_a \neq 0$.

We set the lattice spacing $a=1$ and restrict wavenumbers to the reduced Brillouin zone (BZ) $k \in [-\pi/2,\pi/2]$.
The band index $\alpha=1,2$ serves to map the Fourier-transformed JW fermions from the $2 \pi$  BZ onto the reduced zone as
\begin{equation}
\label{c12}
  c(k)= c_1(k) \cdot \vartheta( \pi/2- |k|)+c_2(k-\pi) \cdot \vartheta(|k|- \pi/2)~,
\end{equation}
where $\vartheta(x)$ is the Heaviside step function. Then the coordinate representation of the JW fermion reads as
\begin{equation}
\label{cFourier}
  c_n=  \frac{1}{\sqrt{N}} \sum_{\alpha,q} c_\alpha(q) (-1)^{(\alpha-1)n} e^{-iqn} ~.
\end{equation}
The Hamiltonian \eqref{FermHam} at $\Delta=0$ can be written as
\begin{equation}
\label{Hspinor}
 H= \frac12 \sum_{k}\psi^{\dag}_{k}\mathcal{H}(k) \psi_{k}~,
\end{equation}
where the fermions are unified in the spinor
\begin{equation}
  \psi_{k}^{\dag}=\left(c_1^{\dag}(k),
  c_2^{\dag}(k),c_1(-k), c_2(-k)\right)~,
\label{spinor1}
\end{equation}
with the $4\times 4$ Hamiltonian  matrix
\begin{equation}
\label{Hk}
  \mathcal{H}(k) = \left(%
\begin{array}{cc}
  \hat{A} & \hat{B} \\
  \hat{B}^\dag  & -\hat{A} \\
\end{array}%
\right)~,
\end{equation}
where
\begin{equation}
\label{A}
  \hat{A} \equiv  \left(%
\begin{array}{cc}
  h +t \cos k & h_a+i \delta \sin k \\
  h_a-i \delta \sin k & h -t \cos k \\
\end{array}%
\right)~,
\end{equation}
and
\begin{equation}
\label{B}
 \hat{B} \equiv  \left(%
\begin{array}{cc}
  -i \gamma \sin k &  -\gamma_a \cos k\\
  \gamma_a \cos k & i \gamma \sin k \\
\end{array}%
\right)~,
\end{equation}
The Hamiltonian has four eigenvalues \cite{Perk:1975} $\pm E_{\pm}$, where
\begin{equation}
\label{Epm}
 E_{\pm}(k)=\sqrt{\mathfrak{C}_2(k)\pm \sqrt{\mathfrak{C}_2^2(k)-\mathfrak{C}_4(k)}}~,
\end{equation}
with
%\begin{widetext}
\begin{equation}
  \label{C2}
  \mathfrak{C}_2(k) \equiv h^2+h_a^2+(t^2+\gamma_a^2) \cos^2k+(\delta^2+\gamma^2)\sin^2k
\end{equation}
and
\begin{widetext}
\begin{equation}
  \label{C4}
  \mathfrak{C}_4(k) \equiv \Big(h^2-h_a^2-(t^2-\gamma_a^2)\cos^2k-(\delta^2-\gamma^2)\sin^2k \Big)^2 +(t \gamma- \delta \gamma_a)^2 \sin^2 2k
\end{equation}
\end{widetext}

\begin{figure}[]
\centering{\includegraphics[width=7.5cm]{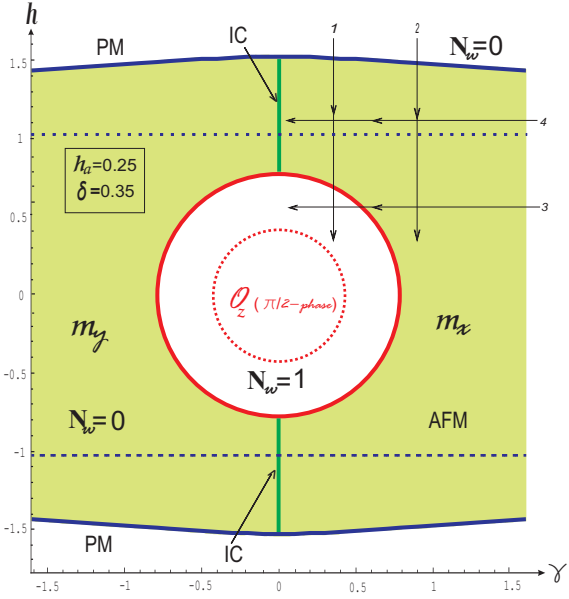}}
\caption{Phase diagram of the model in $h-\gamma$ plane ($\gamma_a=0$). The model is critical  on: \textit{(i)} two infinite
lines $h= \pm h_c^{\s (1)}$ (bold blue);
\textit{(ii)}  circle $h^2+\gamma^2 =\mathcal{R}^2$ (bold red);
\textit{(iii)} two segments $h_c^{\s (2)} \leq |h| \leq h_c^{\s (1)}$ along  $\gamma= 0$ (bold green).
Three phases are shown: disordered paramagnetic (PM) polarized by the axial field, planar antiferromagnetic (AFM) with local order parameters $m_{x,y}$,
and topological $\mathcal{O}_z(\pi/2)$ with oscillating string order. The four paths (1-4) in parametric space are indicated by thin lines. The winding numbers $N_w$ calculated in Sec.~\ref{Res} are also shown. The bold phase boundaries are calculated for interaction $\Delta=1/2$, while their dashed counterparts correspond to non-interacting case $\Delta=0$.}
\label{PDiag}
\end{figure}

The phase diagram of the model \cite{Perk:1975} shown in Fig.\ref{PDiag}, is found from the condition
\begin{equation}
  \label{C4QCP}
  \mathfrak{C}_4(k)=0
\end{equation}
for the critical lines, where the model becomes gapless.

There are three phase boundaries:\\
\textit{(i)} at $\pm h_c^{\s (1)}$ with
\begin{equation}
 \label{Hc}
 h_c^{\s (1)} \equiv \sqrt{t^2+h_a^2-\gamma_a^2}~,~ \forall~ \gamma, \delta
\end{equation}
the gap vanishes at the center of the BZ ($k=0$). \\
\textit{(ii)} At the edge of the BZ ($k= \pm \pi/2$) the gap vanishes on the circle
\begin{equation}
 \label{Circle}
  h^2+\gamma^2=h_a^2+\delta^2~,
\end{equation}
which we will associate with the critical field $h_c^{\s (2)}$.

\textit{(iii)} Two critical line segments at $\gamma=0$ ($\gamma_a=0$) correspond to the gap vanishing at the incommensurate (IC)
wavevector
\begin{equation}
 \label{kIC}
  k_{\s F}= \pm \arcsin Q,~~
Q\equiv   \sqrt{\frac{t^2+h_a^2-h^2}{t^2-\delta^2}}~,
\end{equation}
which corresponds to the Fermi momentum ($\hbar=1$) of the JW fermions.
The IC solution exists in the range of parameters $\gamma=\gamma_a=0$, $|\delta| < 1$, and
\begin{equation}
 \label{ICrange}
\sqrt{h_a^2+\delta^2} \leq |h| \leq \sqrt{t^2+h_a^2}~.
\end{equation}
The Fermi momentum \eqref{kIC} varies continuously from $k_{\s F}=0$ at the intersection of $\gamma=0$ and $h=\pm\sqrt{t^2+h_a^2}$,
to $k_{\s F}= \pm \pi/2$ where the critical segments end at the intersections with the circle.

%
%
%%%%%%%%%%%%%%%%%%%%%%%%%%%%%%%%%%%%%%%%%%%%%%%%%%%%%%%%%%%%%%%%%%%%%%%%%%%%%%
\subsection{Spin and Majorana averages }\label{Aver}
%%%%%%%%%%%%%%%%%%%%%%%%%%%%%%%%%%%%%%%%%%%%%%%%%%%%%%%%%%%%%%%%%%%%%%%%%%%%%%
%
%
%
Differentiation of the free energy with respect to $h$ and to $h_a$ yields two magnetizations
\begin{equation}
\label{mz}
 m_z = \frac1N \sum_{n=1}^{N} \langle \sigma_{n}^{z} \rangle
\end{equation}
and
\begin{equation}
\label{mza}
 m_z^a = \frac1N \sum_{n=1}^{N}(-1)^n \langle \sigma_{n}^{z} \rangle~,
\end{equation}
respectively. Their explicit expressions are:
\begin{widetext}
\begin{equation}
\label{mz1}
m_z= \frac{h}{\pi} \int_{0}^{\pi/2} \left\{\Big(\frac{1}{E_{+}}+\frac{1}{E_{-}}\Big)+
\frac{t^2 \cos^2 k+|w_a|^2 }{R}
\Big(\frac{1}{E_{+}}-\frac{1}{E_{-}}\Big)\right\}dk
\end{equation}
and
\begin{equation}
\label{mza1}
m_z^a= \frac{h_a}{\pi} \int_{0}^{\pi/2} \left\{ \Big(\frac{1}{E_{+}}+\frac{1}{E_{-}}\Big)+
\frac{ |w|^2+ \gamma_a^2 \cos^2 k }{R}
\Big(\frac{1}{E_{+}}-\frac{1}{E_{-}}\Big)\right\}dk~.
\end{equation}
\end{widetext}

We define the auxiliary parameters:
\begin{eqnarray}
\label{zza}
   w &\equiv& h+i\gamma\sin k \\
  w_a &\equiv& h_a+i\delta\sin k \\
  z &\equiv& w w_a -t \gamma_a \cos^2 k \\
  c &\equiv& (ht + h_a \gamma_a) \cos k \\
  R &\equiv& \sqrt{c^2+|z|^2}
\end{eqnarray}

The Hamiltonian \eqref{Hk} is diagonalized with the help of two
unitary $2 \times 2$ matrices $\hat \Phi$ and $\hat \Psi$.
We find
\begin{equation}
\label{Phi}
   \hat \Phi(q)= \left(
                   \begin{array}{cc}
                    e^{-i \theta} \beta_+ & \beta_- \\
                     -\beta_- & e^{i \theta} \beta_+ \\
                   \end{array}
                 \right)~,
\end{equation}
where
\begin{equation}
\label{theta}
  e^{i \theta}  \equiv \frac{z}{|z|} ~,
\end{equation}
and
\begin{equation}
\label{betapm}
  \beta_\pm \equiv \frac{1}{\sqrt{2}} \Big(  1 \pm \frac{c}{R} \Big)^{1/2}~.
\end{equation}
The second matrix of this Bogoliubov transformation is calculated as
\begin{equation}
\label{Psi}
  \hat \Psi =  \hat{I}_E^{-1} \hat \Phi (\hat{A} - \hat{B})~,
\end{equation}
where $\hat{I}_E \equiv \mathrm{diag}(E_+, E_-)$.  We introduce the Majorana fermions as
\begin{equation}
\label{Maj}
   a_n +i b_n  \equiv 2 c^{\dag}_n~.
\end{equation}
From the matrix
\begin{equation}
\label{G}
  \hat{G} (q)  \equiv  \hat{\Psi}^\dag (q) \hat{\Phi} (q)
\end{equation}
we find the correlation function of the Majorana operators:
\begin{equation}
\label{abS}
  \langle ib_n a_m \rangle = \frac{1}{2 \pi}    \int_{-\pi}^{\pi}  dq e^{-iq(m-n)}
  \Big\{ G_{11}(q)+ (-1)^n G_{12}(q)  \Big\}~,
\end{equation}
where the matrix elements are:
\begin{widetext}
\begin{eqnarray}
  G_{11}(q) &=& (t \cos q +w^\ast) \Big\{ \frac{\beta_+^2}{E_+}+ \frac{\beta_-^2}{E_-} \Big\} +
   (w_a-\gamma_a \cos q)  e^{-i \theta} \beta_+ \beta_-
   \Big\{ \frac{1}{E_+}- \frac{1}{E_-} \Big\}~,  \label{G11} \\
  G_{12}(q) &=& (w_a-\gamma_a \cos q) \Big\{ \frac{\beta_-^2}{E_+}+ \frac{\beta_+^2}{E_-} \Big\}+
    (t \cos q +w^\ast) e^{i \theta} \beta_+ \beta_-
     \Big\{ \frac{1}{E_+}- \frac{1}{E_-} \Big\}~.  \label{G12}
\end{eqnarray}
\end{widetext}
All the above formulas recover those of \cite{Chitov:2019} in the limit $t \to 1$ and $\gamma_a \to 0$.
The correlation function \eqref{abS} is a building element of Toeplitz determinants \cite{LiebSM:1961} used to
calculate local order parameters (magnetization) and nonlocal SOPs. Addition of $\gamma_a \neq 0$
only slightly numerically modifies positions of boundaries on the phase diagram and the values of correlation
functions, leaving the structure of the phase diagram, the nature of its phases, and order parameters essentially the same
as reported in our earlier work \cite{Chitov:2019}, see Fig.\ref{PDiag}.

%
%
%
%xxxxxxxxxxxxxxxxxxxxxxxxxxxxxxxxxxxxxxxxxxxxxxxxxxxxxxxxxxxxxxxxxxxxxxxxxxxxxx
%
\section{Mean-field equations}\label{MFA}
%
%xxxxxxxxxxxxxxxxxxxxxxxxxxxxxxxxxxxxxxxxxxxxxxxxxxxxxxxxxxxxxxxxxxxxxxxxxxxxxx
%
%
%
The mean-field theory for the $XYZ$ chain is in fact the Hartee-Fock approximation for its interacting fermionic
representation \eqref{FermHam}. We use the most general decoupling \cite{Santos:1989} for the interacting term with a product of two number operators
($\hat n_l=c_l^\dag c_l$) as
\begin{eqnarray}
\label{HFA}
  \hat n_l \hat n_m &\approx&  \hat n_l \langle \hat n_m\rangle + \hat n_m  \langle \hat n_l \rangle -
      \langle \hat n_l\rangle  \langle \hat n_m \rangle \nonumber \\
  &+& c_l^\dag c_m \langle  c_l c_m^\dag  \rangle   + \mathrm{H.c.} + |\langle  c_l c_m^\dag  \rangle |^2  \nonumber \\
  &+& c_l^\dag c_m^\dag  \langle  c_m c_l  \rangle   + \mathrm{H.c.} - |\langle  c_l c_m  \rangle |^2
\end{eqnarray}
Such approximation applied to the Heisenberg chain is known from the literature to be accurate, at least qualitatively, see, e.g., \cite{Santos:1989,Dmitriev:2002JETP,*Dmitriev:2002PRB,Caux:2003,*Caux:2005,Yamamoto:2000}. One cannot expect the mean-field approximation to furnish, e.g.,
correct critical indices to identify the universality class, but predictions of model's phase diagram and order parameters are qualitatively correct. Since $1d$ is a realm of strong fluctuations, special care needs to be exercised while dealing with the mean-field predictions for phase boundaries (critical points). They need to be cross-checked against available exact results, as we will explain below.

We introduce the following mean-field parameters:
\begin{eqnarray}
\label{teta}
  \langle  c_n c_{n+1}^\dag \rangle &\equiv&   \mathcal{K}+(-1)^n \delta\eta \\
\label{Peta}
   \langle  c_n c_{n+1} \rangle &\equiv& P- (-1)^n \delta\eta_{\s P} \\
\label{MMa}
  \langle 1- 2 c_n^\dag c_n  \rangle &\equiv& m_z+(-1)^n m_z^a
\end{eqnarray}
Using decoupling \eqref{HFA} and parameters \eqref{teta}-\eqref{MMa} in  \eqref{FermHam}, we obtain the
approximate mean-field Hamiltonian
\begin{equation}
\label{HMF}
 H \approx H_{\s MF}=N \Delta \mathcal{C}+  \frac12 \sum_{k}\psi^{\dag}_{k}\mathcal{H}_{\s R}(k) \psi_{k}~.
\end{equation}
The renormalized Hamiltonian $\mathcal{H}_{\s R}(k)$ is given by the same expressions as for the non-interacting case \eqref{Hk}, \eqref{A}, and \eqref{B}, with the difference that the six bare couplings of the free-fermionic Hamiltonian are replaced by the remormalized parameters
as follows:
\begin{eqnarray}
\label{hR}
    h &\longmapsto&  h_{\s R} \equiv h- \Delta m_z \\
\label{haR}
    h_a &\longmapsto& h_{a \s R} \equiv h_a+ \Delta m_z^a \\
\label{tR}
    t &\longmapsto&  t_{\s R} \equiv 1+2\Delta(\mathcal{K} +\delta^2\eta) \\
\label{dR}
    \delta &\longmapsto& \delta_{\s R} \equiv \delta \big(1+2\Delta(\mathcal{K} +\eta) \big) \\
\label{gR}
    \gamma  &\longmapsto& \gamma_{\s R} \equiv \gamma-2\Delta(P-\delta^2\eta_{\s P})\\
\label{gaR}
    \gamma_a &\longmapsto& \gamma_{a \s R} \equiv \gamma_a- 2\Delta\delta(P-\eta_{\s P})
\end{eqnarray}
and the constant term is
\begin{equation}
\label{Const}
\mathcal{C}= \mathcal{K}^2-P^2-\frac14 m_z^2+ \frac14 (m_z^a)^2 +
\delta^2 \big( \eta^2-\eta_{\s P}^2  +2  \mathcal{K} \eta+ 2P\eta_{\s P} \big)
\end{equation}
Contrary to the model's bare parameters of choice, the renormalized couplings \eqref{hR}-\eqref{gaR} are to be found from a set of
six self-consistent equations obtained from minimization of the free energy. The latter is calculated from the Hartree-Fock Hamiltonian
$\mathcal{H}_{\s R}(k)$. Using \eqref{teta} we find equations for the bond average
\begin{widetext}
\begin{eqnarray}
\label{t}
  \mathcal{K} = \frac{ t_{\s R}}{2\pi}\int_{0}^{\pi/2} dk \cos^2k \Big\{
   \frac{1}{E_+}+ \frac{1}{E_-}+ \frac{h_{\s R}^2+\gamma^2_{a \s R}\cos^2k}{R} \Big(\frac{1}{E_+}-\frac{1}{E_-}\Big)
  +  \frac{\delta_{\s R}\gamma_{\s R} \gamma_{a \s R}\sin^2 k}{ t_{\s R} R} \Big(\frac{1}{E_+}- \frac{1}{E_-}\Big) \Big\}
\end{eqnarray}
and for the dimerization susceptibility $\eta$
\begin{eqnarray}
\label{eta}
 \delta  \eta = \frac{\delta_{\s R}}{2\pi}\int_{0}^{\pi/2} dk \sin^2k \Big\{
   \frac{1}{E_+}+ \frac{1}{E_-}+ \frac{|w|^2}{R} \Big(\frac{1}{E_+}-\frac{1}{E_-}\Big)
  +  \frac{t_{\s R}\gamma_{\s R} \gamma_{a \s R}\cos^2 k}{\delta_{\s R} R} \Big(\frac{1}{E_+}- \frac{1}{E_-}\Big) \Big\}~.
\end{eqnarray}
From \eqref{Peta} we obtain equations for the anomalous pairing amplitude
\begin{eqnarray}
\label{P}
 P = \frac{\gamma_{\s R}}{2\pi}\int_{0}^{\pi/2} dk \sin^2k \Big\{
   \frac{1}{E_+}+ \frac{1}{E_-}+ \frac{|w_a|^2}{R} \Big(\frac{1}{E_+}-\frac{1}{E_-}\Big)
  +  \frac{t_{\s R} \delta_{\s R} \gamma_{a \s R}\cos^2 k}{ \gamma_{\s R} R} \Big(\frac{1}{E_+}- \frac{1}{E_-}\Big) \Big\}
\end{eqnarray}
and for the anomalous susceptibility
\begin{eqnarray}
\label{etaP}
 \delta  \eta_{\s P} = -\frac{ \gamma_{a \s R}}{2\pi}\int_{0}^{\pi/2} dk \cos^2k \Big\{
   \frac{1}{E_+}+ \frac{1}{E_-}+ \frac{h_{a \s R}^2+t^2_{ \s R}\cos^2k}{R} \Big(\frac{1}{E_+}-\frac{1}{E_-}\Big)
  +  \frac{ t_{\s R} \delta_{\s R}\gamma_{\s R} \sin^2 k}{\gamma_{a \s R} R} \Big(\frac{1}{E_+}- \frac{1}{E_-}\Big) \Big\}~.
\end{eqnarray}
\end{widetext}
The uniform and staggered magnetizations  \eqref{MMa} satisfy equations \eqref{mz1} and \eqref{mza1}
with their right hand sides written in terms of the renormalized couplings \eqref{hR}-\eqref{gaR}.
In the following we chose the bare coupling $\gamma_a=0$.

\begin{figure}[]
\centering{\includegraphics[width=8.5cm]{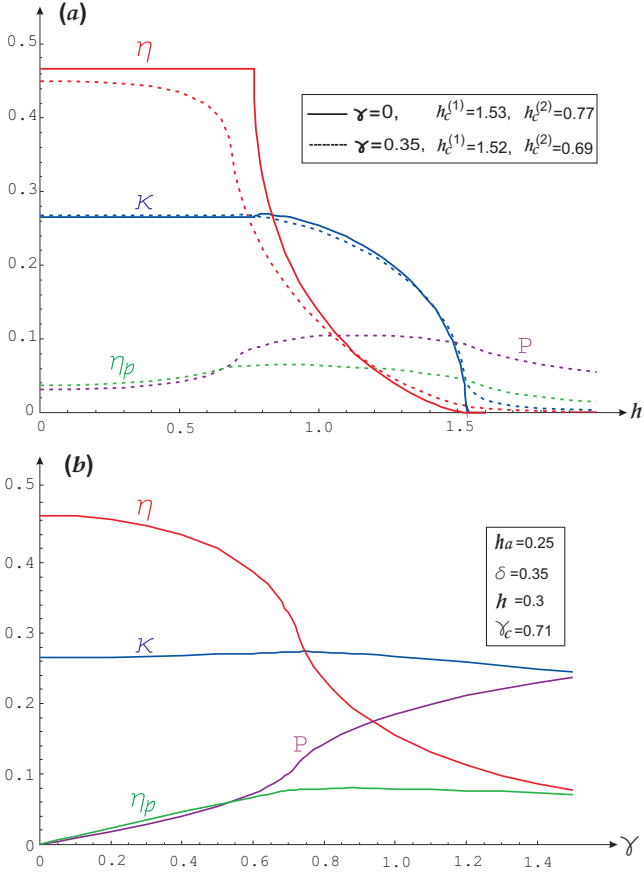}}
        \caption{Representative behavior of mean-field parameters \eqref{t}-\eqref{etaP} calculated for interaction $\Delta=1/2$. The panel (a) shows results along the path 1 on the phase diagram in Fig.~\ref{PDiag}. The panel (b) corresponds to path 3. In addition, panel (a) presents results along the line $\gamma=0$.}
\label{MFPars}
\end{figure}

The mean-field parameters \eqref{t}-\eqref{etaP} are fundamentally important for calculation of the phase diagram, the local and string
order parameters in different phases. The representative numerical results for these parameters are shown in Fig.~\ref{MFPars}. Note that anomalous average $P$ and $\eta_{\s P}$ are not the true (superconducting) order parameters
signalling spontaneous breaking of $U(1)$ symmetry. This symmetry is intrinsically broken by model's anisotropy couplings $\gamma, \gamma_a$.
As one can see from Fig.~\ref{MFPars}(b), in the symmetry-restoring  limit  $\gamma, \gamma_a \to 0$, the anomalous average parameters vanish.

%
%
%

%
%
%
%xxxxxxxxxxxxxxxxxxxxxxxxxxxxxxxxxxxxxxxxxxxxxxxxxxxxxxxxxxxxxxxxxxxxxxxxxxxxxx
%
\section{Results for $XYZ$ chain}\label{Res}
%
%xxxxxxxxxxxxxxxxxxxxxxxxxxxxxxxxxxxxxxxxxxxxxxxxxxxxxxxxxxxxxxxxxxxxxxxxxxxxxx
%
%
%
Before we proceed to explore predictions of the derived mean-field equations, let us first understand qualitatively possible outcomes.
The way the mean-field theory is constructed, i.e., by switching to the renormalized couplings \eqref{hR}-\eqref{gaR}, makes it obvious
that the interacting model has the same spectrum  as in Eq.~\eqref{Epm}, but with renormalized parameters. Thus we obtain the same phases and their order parameters, conditions
for the phase boundaries (gaplessness), etc, as described above for the case $\Delta=0$ (see \cite{Chitov:2019} for more details), proviso
that all bare couplings are renormalized in appropriate formulas. Within present theory, no new phase with a new order parameter, other than
presented on the phase diagram in Fig.~\ref{PDiag}, can occur.

Interactions, however, can bring about additional nontrivial solutions of the mean-field equations for the renormalized parameters, like, dimerization, anisotropy, uniform or staggered fields/magnetizations, even when their bare counterparts are zero. That would constitute the case of spontaneous symmetry breaking associated with a phase transition. As one can see from Fig.~\ref{MFPars}b, the anomalous average parameters vanish in the limit $XYZ \to XXZ$. We did not find numerical signs of spontaneous breaking of the $U(1)$ symmetry (superconductivity) at $\Delta \neq 0$. Neither we found spontaneous dimerization when bare $\delta=0$. This is in agreement with available results for the $XYZ$ and $XXZ$ models \cite{Takahashi:1999,Franchini:2017,denNijs:1981,Alcaraz:1995}. However, it is known from exact results that $\Delta = \pm 1$ are critical points of the antiferro-/ferromagnetic phase transitions in the $XXZ$ model \cite{Takahashi:1999,Franchini:2017}. In the $XYZ$ chain ($\gamma \neq 0$) spontaneous antiferro-/ferromagnetism appears at $|\Delta|>1$
\cite{denNijs:1981}. To stay on the safe side and to avoid dealing with the interaction-induced magnetism in the results which follow, we will assume the regime of weak interaction $|\Delta| <1$ in this section. The srongly-interacting regime $\Delta \gtrsim 1$ is discussed in Sec.~\ref{XXZRes} for the $XXZ$ chain.

The phase diagram of the model is shown in Fig.~\ref{PDiag}.
Overall, the mean-field results in this regime are qualitatively similar to the non-interacting ($\Delta =0$) case \cite{Chitov:2019}.
The PM-AFM boundary \eqref{Hc} gets modified by interactions. It is not a straight line anymore. The value for critical field $h_c^{(1)}$ is
available only numerically. However its maximum value reached in the $XXZ$ limit is found exactly from our equations:
\begin{equation}
\label{hc1XXZ}
  \gamma=0:~h_c^{(1)}=\Delta +\sqrt{1+h_a^2}~,
\end{equation}
in agreement with earlier scaling results \cite{Alcaraz:1995}.

\begin{figure}[]
\centering{\includegraphics[width=7.5cm]{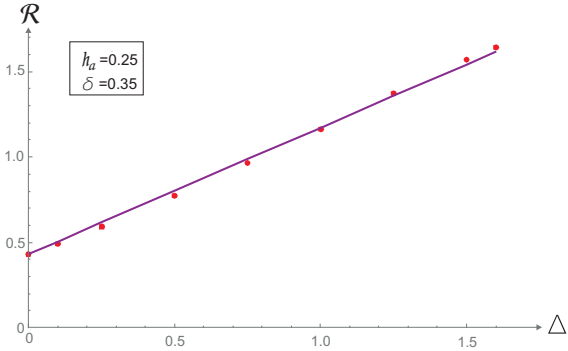}}
\caption{Radius of the circle enclosing  $\mathcal{O}_z(\pi /2)$-phase in the phase diagram Fig.~\ref{PDiag} as a function of interaction. The linear fit with $a \approx 0.745$ is shown. }
\label{RDelta}
\end{figure}

The topological phase with oscillating string order is located inside the circle on the phase diagram in Fig.~\ref{PDiag}. Quite amazingly (in the view of complexity of the six coupled mean-field equations), interactions only change the radius of the circle $\mathcal{R}$, conserving the perfect shape of this phase boundary. Numerically we found
\begin{equation}
\label{R}
  \mathcal{R}(\Delta) \approx  \mathcal{R}(0) +  a \Delta~,
\end{equation}
where  the radius for the non-interacting case $\mathcal{R}(0)=\sqrt{h_a^2+\delta^2}$.
The linear fit with $a \approx 0.745$, shown in Fig~\ref{RDelta}, works quite well even at $\Delta \gtrsim 1$.
%
%
%

%
%
%%%%%%%%%%%%%%%%%%%%%%%%%%%%%%%%%%%%%%%%%%%%%%%%%%%%%%%%%%%%%%%%%%%%%%%%%%%%%%
\subsection{Induced and spontaneous magnetizations}\label{Magn}
\begin{figure}[]
\centering{\includegraphics[width=8.5cm]{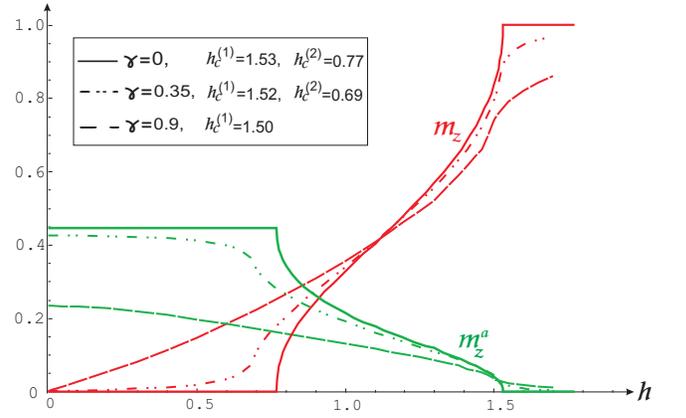}}
        \caption{Two induced magnetizations $m_z$ (red) and $m_z^a$ (green) \textit{vs} uniform field $h$ at  $\delta=0.35$,  $h_a=0.25$, $\Delta=0.5$
        for different $\gamma$. At $\gamma=0$, $m_z$ (solid line) demonstrates plateaux in the gapped phases connected by a continuous curve through the gaplees IC phase. Similar behavior is demonstrated by $m_z^a$. Dashed-dotted lines correspond to path 1 shown in the phase diagram Fig.~\ref{PDiag}. Dashed lines correspond to path 2.  The magnetizations show noticeable cusps at the critical fields $h_c^{(1)}$ (path $2$); $h_c^{(1)}$ and $h_c^{(2)}$ (paths $\gamma =0$ and $1$), when the paths cross phase boundaries.}
\label{Mzall}
\end{figure}

%%%%%%%%%%%%%%%%%%%%%%%%%%%%%%%%%%%%%%%%%%%%%%%%%%%%%%%%%%%%%%%%%%%%%%%%%%%%%%
%
%
%
First we present the field-induced magnetizations  $m_z$ and $m_z^a$ as functions of the uniform magnetic field $h$ in Fig.~\ref{Mzall}.
Their explicit expressions \eqref{mz1} and \eqref{mza1} are calculated at each point with the renormalized couplings on the
right hand sides, determined self-consistently from numerical solution of the mean-field equations given in the previous section.

The plots for the $XYZ$ chain are done for two cases. The first case corresponds to the path $1$ on the phase diagram in the $h-\gamma$ plane shown  in Fig.~\ref{PDiag}. The path crosses the PM-AFM boundary at $h=h_c^{\s (1)}$ and the AFM-$\mathcal{O}_z(\pi/2)$ boundary  at $h=h_c^{\s (2)}$. The magnetizations have noticeable cusps at these critical points, which correspond to divergent susceptibilities.
In case of the path $2$, it crosses only the PM-AFM boundary and bypasses the topological phase. The magnetizations demonstrate cusps at the only critical point $h_c^{\s (1)}$, while at $h< h_c^{\s (1)}$ they and their derivatives are analytical.

The phase diagram in Fig.~\ref{PDiag} contains two conventional antiferromagnetic phases with spontaneous planar magnetizations
$m_x$ and $m_y$. The local order parameter $m_x$ is calculated form the limit of  the spin-correlation function which is also the
correlation function of the Majorana string operators \cite{LiebSM:1961}:
\begin{equation}
\label{Mx}
  \langle \sigma_L^x \sigma_R^x \rangle =
 \Big  \langle  \prod_{n=L}^{R-1} \big[ i b_n a_{n+1} \big] \Big \rangle
   \xLongrightarrow[R \to \infty]{~} m_x^2~.
\end{equation}
As we have shown in \cite{Chitov:2019}, this  Majorana string correlation function is given by the determinant of the block Toeplitz matrix constructed from $\frac12 (R-L)\times \frac12 (R-L)$ blocks of size $2 \times 2$ with the elements given by Eq.~\eqref{abS}. For explicit expressions of this block
Toeplitz matrix we refer the reader to \cite{Chitov:2019}. At each point in the parametric space the elements \eqref{abS} of this matrix  are calculated with renormalized couplings determined from the mean-field equations. The results for spontaneous magnetization are given in
Fig.~\ref{OPsall}. The numerical values of the parameters we present in that figure are stable in the fourth decimal place for the
$M \times M$ matrices of  sizes $M \gtrsim 30$. In immediate vicinities of the critical points the order parameters are checked to decay smoothly as
$M \to \infty$. The expressions for $m_y$ are obtained along the same lines. Numerical values satisfy useful relation $m_y(-\gamma)=m_x(\gamma)$,
verified explicitly.

%
%
%%%%%%%%%%%%%%%%%%%%%%%%%%%%%%%%%%%%%%%%%%%%%%%%%%%%%%%%%%%%%%%%%%%%%%%%%%%%%%
\subsection{Nonlocal string order}\label{String}
%%%%%%%%%%%%%%%%%%%%%%%%%%%%%%%%%%%%%%%%%%%%%%%%%%%%%%%%%%%%%%%%%%%%%%%%%%%%%%
%
%
%
Now we address the topological phase with nonlocal string order inside the circle  in Fig.~\ref{PDiag},
first reported in \cite{Chitov:2019} for non-interacting case. It turns out that the fermionic interaction
renormalizes the phase boundary and SOPs, but does not alter the nature of the order in this phase.
To quantify this type of order we use the string operator
\begin{equation}
\label{Oz}
  O_z(n) \equiv \prod_{l=1}^{n} \sigma_l^z = \prod_{l=1}^{n} \big[ i b_{l} a_{l} \big]~,
\end{equation}
and related string correlation function
\begin{equation}
\label{Dzz}
  \mathfrak{D}_{zz}(L,R) \equiv
 \langle O_z(L-1) O_z(R) \rangle =
 \Big \langle  \prod_{l=L}^{R} \big[ i b_l a_l \big]  \Big \rangle~.
\end{equation}

Following the original proposal by den Nijs and Rommelse \cite{denNijs:1989}, the SOP
was defined and detected in the subsequent work on the spin chains, see, e.g., \cite{Hida:1992,*Hida:1992b,Hatsugai:1991,Kennedy:1992,*Kohmoto:1992,*Oshikawa:1992}.
The SOP was defined (up to some minor variations) as the limit of the string-string  correlation function,
which is not convenient, since such SOP has a wrong dimension of square of the order parameter.
The definition we use, due to Berg \textit{et al} \cite{Berg:2008}, is more consistent with the standard theory of critical phenomena,
and the critical index of the (string) order parameter $\beta$ correctly enters all the hyperscaling relations
\cite{Chitov:2017JSM}.
The correlation function  \eqref{Dzz} is calculated from the determinant of the block Toeplitz matrices,
built from elements \eqref{abS}. These matrices are given explicitly in \cite{Chitov:2019}.
\begin{figure}[]
\centering{\includegraphics[width=7.5cm]{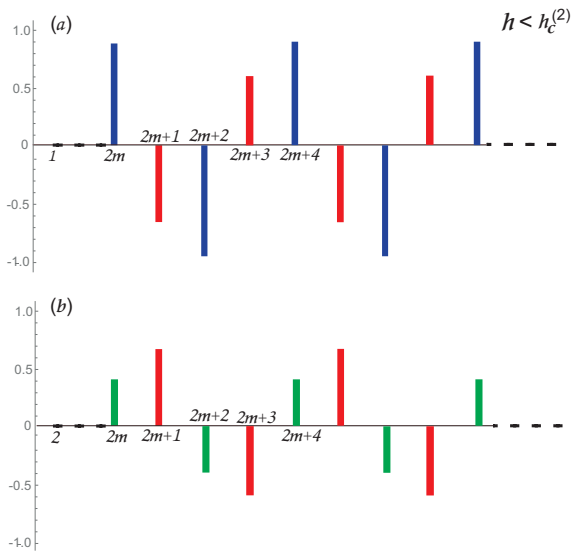}}
\caption{Visualization of the oscillating string order inside the circle at the point $h=0.2$ on path 1 (Fig.~\ref{PDiag}) for
$h_a=0.25$, $\delta=0.35$, $\gamma=0.35$, and $\Delta=1/2$. Panel (a) shows $\mathfrak{D}_{zz}(1,N)$ \eqref{Ozeo} with alternating limiting values $\pm \mathcal{O}_{z,1}^2$ (blue) and $\pm \mathcal{O}_{z,3}^2$ (red). Panel (b) shows $\mathfrak{D}_{zz}(2,N)$ with similar parameters $\pm \mathcal{O}_{z,2}^2$ (green) and $\pm \mathcal{O}_{z,3}^2$.}
\label{SOPDetails}
\end{figure}

Inside the circle, $ \mathfrak{D}_{zz}$ oscillates with the period of four lattice spacings (i.e., \textit{twice the unit cell}), see Fig.~\ref{SOPDetails}. Doubling of the translational period by the string order is a sign of spontaneous breaking of the hidden
$\mathbb{Z}_2 \otimes \mathbb{Z}_2$  symmetry.
This phase is labeled as $\mathcal{O}_z(\pi /2)$ to distinguish it from the plain behavior of $ \mathfrak{D}_{zz}$ in the PM phase.
Since $\mathfrak{D}_{zz}(L,R) \neq \mathfrak{D}_{zz}(R-L)$, we need three parameters to account for the string order:
\begin{widetext}
\begin{equation}
\label{Ozeo}
 \mathfrak{D}_{zz}(L,R)
 \xrightarrow[R \to \infty]{~}
  \left\{
    \begin{array}{c}
      (-1)^m \mathcal{O}_{z,1}^2~,~~~~~ L=1,~R=2m \\[0.2cm]
      (-1)^m \mathcal{O}_{z,2}^2~,~~~~ L=2,~R=2m \\[0.2cm]
      (-1)^{m+L} \mathcal{O}_{z,3}^2~,~~~~~ L=1,~R=2m+1~ \mathrm{or}~ L=2,~R=2m+1  \\
    \end{array}
   \right.
\end{equation}
%\end{widetext}
The ordering patterns \eqref{Ozeo} detected from non-decaying oscillations of the string correlation function for a particular parametric point in
the $\mathcal{O}_z(\pi /2)$-phase, are depicted in Fig.~\ref{SOPDetails}.
The magnitudes of the SOPs  $\mathcal{O}_{z,i}$ along different paths on the phase diagram Fig.~\ref{PDiag} are given in Fig.~\ref{OPsall}(a)
and (c).  At each point the SOP is calculated with the remormalized couplings determined from numerical solution of the self-consistent mean-field equations. Similarly to the non-interacting case \cite{Chitov:2019}, the other two components of the SOP  $\mathcal{O}_x$ and  $\mathcal{O}_y$
vanish when $h \neq 0$ and $h_a \neq 0$.

The string correlation function $\mathfrak{D}_{zz}$ in the PM saturated phase is always positive and essentially monotonous. For completeness we plot in Fig.~\ref{OPsall} the PM SOP defined as $\lim_{R \to \infty} \mathfrak{D}_{zz}(L,R)=  \mathcal{O}_z^2$.

%\begin{widetext}
\begin{figure}[]
\centering{\includegraphics[width=17cm]{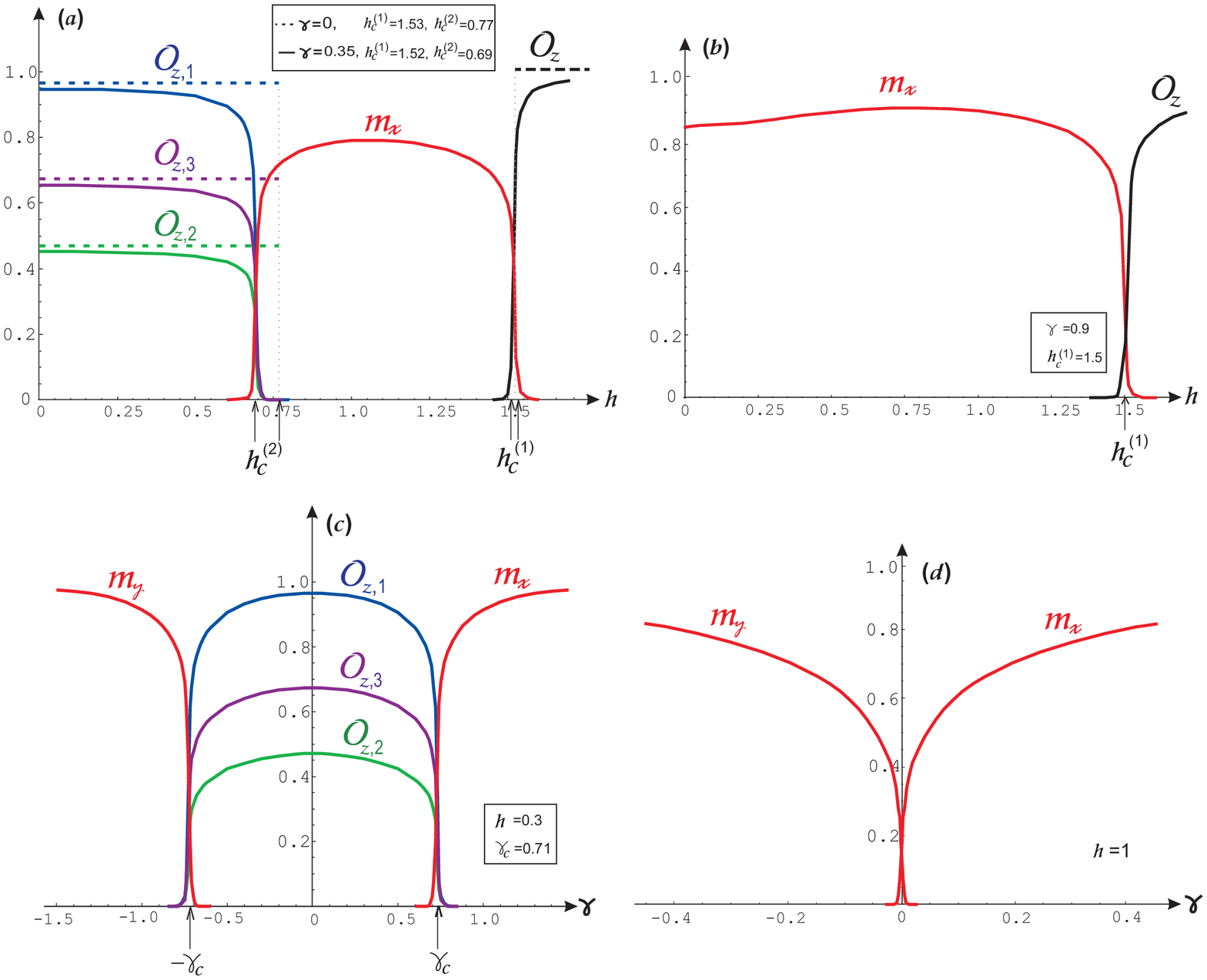}}
\caption{Spontaneous planar magnetizations $m_{x,y}$ and modulated string order parameters $\mathcal{O}_{z,i}$ numerically calculated from the $2N \times 2N$ matrices with $N=70$. The panels (a-d) correspond to the paths 1-4 on the phase diagram shown in Fig.~\ref{PDiag} ($h_a=0.25$ and $\delta=0.35$).
In the AFM phases $m_{x,y} \neq 0$ and $\mathcal{O}_{z,i}=0$. In the PM phase $h>h_c^{(1)}$,  $\mathcal{O}_{z} \neq 0$ and plainly monotonous. The exact values of the critical parameters $h_c^{(1)},h_c^{(2)},\gamma_c$ (shown by arrows) are calculated from renormalized Eqs.~(\ref{Hc},\ref{Circle}). Non-vanishing small tails of the order parameters seen in the immediate vicinities of the critical points are the finite-size effects, checked to die off as $N \to \infty$. The special case $\gamma=0$ when all SOPs become step-like functions is shown in panel (a).}
\label{OPsall}
\end{figure}
\end{widetext}

\begin{figure}[]
\centering{\includegraphics[width=7.5cm]{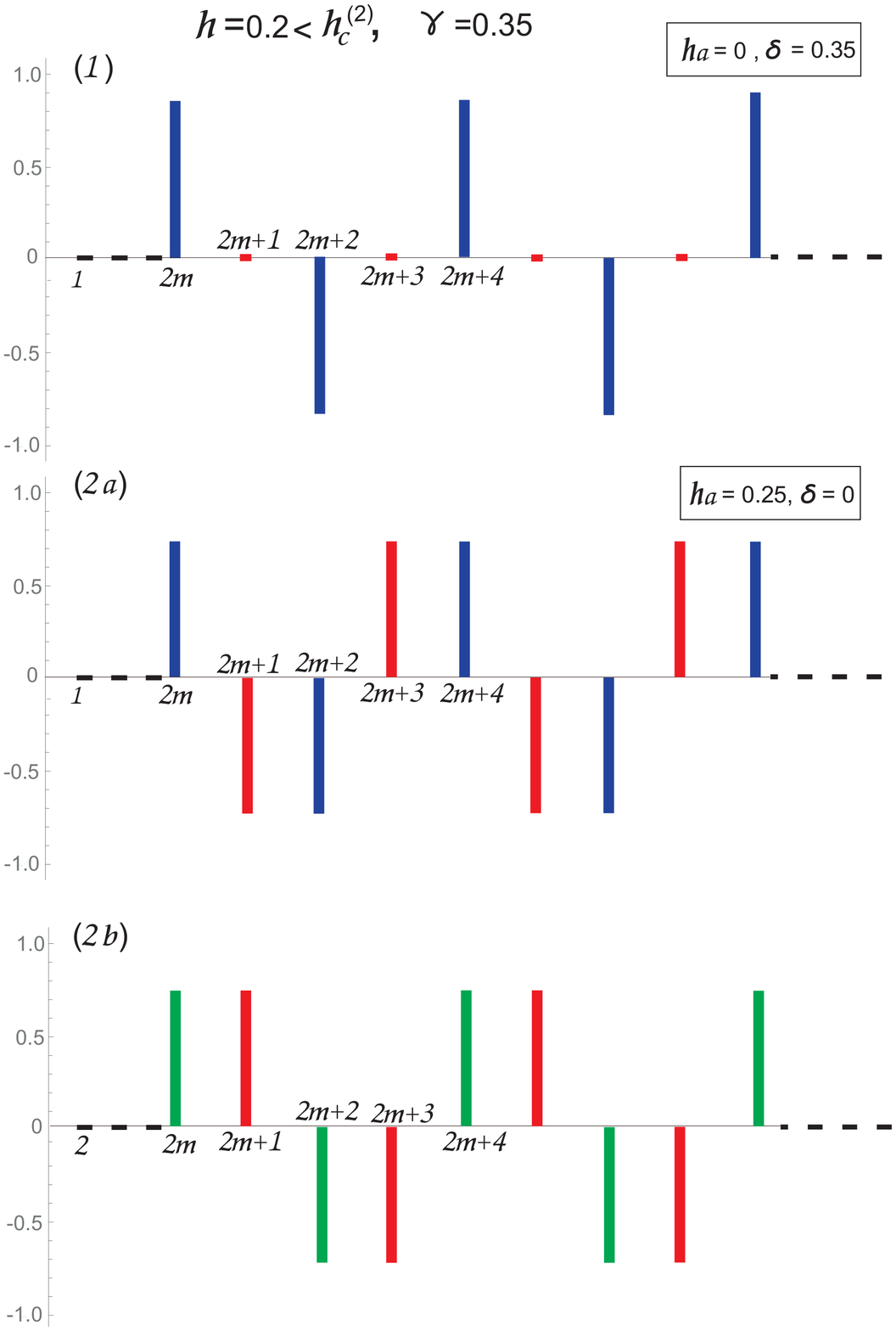}}
\caption{Visualization of the oscillating string order inside the circle at the point $h=0.2$ on path 1 (Fig.~\ref{PDiag}) with
$\gamma=0.35$ and $\Delta=1/2$ for two special cases. (1): For $h_a=0$ and $\delta=0.35$ the string order shown in Fig.~\ref{SOPDetails}(b)
vanishes, since $\mathcal{O}_{z,2}=\mathcal{O}_{z,3}=0$, while the order shown in Fig.~\ref{SOPDetails}(a) reduces to the pattern (1) above.
(2): For  $h_a=0.25$ and $\delta=0$,  $\mathcal{O}_{z,2}=\mathcal{O}_{z,2}=\mathcal{O}_{z,3} \neq 0$, and the patterns shown in Fig.~\ref{SOPDetails}
become the same, up to a singe lattice spacing translation,  as shown in panels (2a) and (2b) above.}
\label{SOPDetailsSplit}
\end{figure}

There are interesting limiting cases of the topological string order.  Two alternating (bare) parameters of the model, $h_a$ and $\delta$  generate the topological phase, see  Fig.~\ref{PDiag}.  The radius of its boundary $\mathcal{R} = \sqrt{h_{a \s R}^2+\delta_{\s R}^2}$. We check from the mean-field equations that at $|\Delta| <1$, $h_{a \s R} \propto h_a$ and $\delta_{\s R} \propto  \delta$, i.e. turning off one of those parameters, turns off its renormalized counterpart as well. Although the four lattice spacing periodicity of the string correlation function \eqref{Ozeo} is preserved, its ordering patterns are distinct. There are often physically interesting situations when there is an alternating component of the magnetic field (or modulated chemical potential, when dealing with various versions of the Kitaev-Majorana models \eqref{FermHam}), while the dimerization is absent. Or vice versa, quite often one is dealing with dimerized models with uniform magnetic field (chemical potential). We find for the former case:
\begin{equation}
\label{ha0}
  h_a=0, ~\delta \neq 0:~~ \mathcal{O}_{z,1} \neq 0,~ \mathcal{O}_{z,2}= \mathcal{O}_{z,3}=0~,
\end{equation}
and for the latter:
\begin{equation}
\label{delta0}
  h_a \neq 0, ~\delta = 0:~~ \mathcal{O}_{z,1}=\mathcal{O}_{z,2}= \mathcal{O}_{z,3}~,
\end{equation}
These properties hold for the non-interacting case ($\Delta =0$) as well as in the presence of interactions ($\Delta \neq 0$).
Two cases of the ordering patterns are shown in Fig.~\ref{SOPDetailsSplit}.

With the help of duality mappings \cite{Chitov:2017JSM} and identities for the string operators \cite{Chitov:2019}, we find the SOP analytically for $\Delta=0$ inside the circle for the case $h_a=0$ along the line $h=0$:
\begin{equation}
\label{Oz1dg}
  h_a=h=0:~~ \mathcal{O}_{z,1}^2=
   2 \left[\frac{(\delta^2- \gamma^2)}{((1 \pm \delta)^2-\gamma^2)^2} \right]^{1/4}.
\end{equation}
The above result yields the critical index of the order parameter $\beta = 1/8$ in the universality class of the 2D Ising model.
Eq.~\eqref{Oz1dg} is derived for $\Delta =0$, the interacting result within the present approximation is obtained by promoting bare couplings in \eqref{Oz1dg}  to the renormalized ones. \footnote{\label{CritInd} It should be kept in mind that the critical indices found for $\Delta=0$
are not valid for the interacting case \cite{Luther:1975,denNijs:1981}}

Note that the case  \eqref{ha0} applies for the dimerized isotropic ($\gamma=0$) Heisenberg chain without magnetic field. In the $SU(2)$ limit
$\Delta=1$ the non-interacting result  \eqref{Oz1dg} can be improved. The magnitude of the SOP  $\mathcal{O}_{z,1}$ was calculated by Hida via
bosonization \cite{Hida:1992,*Hida:1992b}, however the oscillating pattern of the string order shown in Fig.~\ref{SOPDetailsSplit} (1) was not reported before.

To deal with  the string order in a more unified and compact way, we introduce a new function
\begin{equation}
\label{Dzpl}
\mathfrak{D}_{zz}^{(+)}(n) \equiv  \mathfrak{D}_{zz}(1,n) +  \mathfrak{D}_{zz}(2,n)
\end{equation}
From visual inspection of the patterns shown in Fig.~\ref{SOPDetails} (a) and (b) one can easily check that $\mathfrak{D}_{zz}^{(+)}(n)$
has its ordering pattern similar to the one shown in Fig.~\ref{SOPDetailsSplit} (1), i.e.,
\begin{equation}
\label{DzplAn}
\mathfrak{D}_{zz}^{(+)}(n) \xrightarrow[n \to \infty]{~} \cos \Big( \frac{\pi}{2}n \Big) \mathcal{O}_{z,+}^2~,
\end{equation}
where
\begin{equation}
\label{Ozpl}
 \mathcal{O}_{z,+}^2 \equiv  \mathcal{O}_{z,1}^2 + \mathcal{O}_{z,2}^2~.
\end{equation}
From inspection of Fig.~\ref{SOPDetailsSplit} one can check as well that the special cases   \eqref{ha0} and \eqref{delta0} can be  united under the same
pattern of Eq.~\eqref{DzplAn}.

%
%
%%%%%%%%%%%%%%%%%%%%%%%%%%%%%%%%%%%%%%%%%%%%%%%%%%%%%%%%%%%%%%%%%%%%%%%%%%%%%%
\subsection{Winding number}\label{Nw}
%%%%%%%%%%%%%%%%%%%%%%%%%%%%%%%%%%%%%%%%%%%%%%%%%%%%%%%%%%%%%%%%%%%%%%%%%%%%%%
%
%
%
For each phase we also find the winding number. The calculation outlined in \cite{Chitov:2019}, for the quadratic Hamiltonian
\eqref{Hk},\eqref{A}, \eqref{B} leads to the following result:
\begin{equation}
\label{NwAn}
  N_w=\frac{1}{2\pi i} \Big [ \ln \lambda_+(k)+\ln \lambda_-(k)) \Big]_{-\frac{\pi}{2}^{\s +}}^{\frac{\pi}{2}^{\s -}}~,
\end{equation}
where
\begin{widetext}
\begin{equation}
\label{lambdapm}
    \lambda_\pm (k)=h \pm
    \big(h_a^2+(t^2-\gamma_a^2)\cos^2k+(\delta^2-\gamma^2)\sin^2k - i (t \gamma- \delta \gamma_a) \sin 2k \big)^{1/2}
\end{equation}
\end{widetext}
are the eigenvalues of $\hat{D}(k) \equiv \hat{A} (k)+\hat{B}(k)$. One can establish an important relation between parameter $\mathfrak{C}_4$ of the Hamiltonian's spectrum, defined by \eqref{C4},  and eigenvalues \eqref{lambdapm}:
\begin{equation}
\label{C4Lpm}
  \mathfrak{C}_4 =| \lambda_+|^2 | \lambda_-|^2~.
\end{equation}
A simple comparison of the condition \eqref{C4QCP} for quantum criticality and Eq.~\eqref{C4Lpm} leads to the following conclusion: \textit{topological winding number (mod 2) can change only upon crossing gapless phase boundary}. Within the present approach, numbers $N_w$ in different phases of interacting model are calculated using Eqs.~(\ref{NwAn},\ref{lambdapm}) with  renormalized couplings. Their values are shown in Fig.~\ref{PDiag}. Only the phase with oscillating string order is topologically non-trivial, $N_w=1$.

%
%
%
%xxxxxxxxxxxxxxxxxxxxxxxxxxxxxxxxxxxxxxxxxxxxxxxxxxxxxxxxxxxxxxxxxxxxxxxxxxxxxx
%
\section{Isotropic chain}\label{XXZRes}
In this section we present the results for isotropic $XXZ$ chain, that is the limit $\gamma=\gamma_a=0$. It turns out that a considerable progress can be achieved in analytical treatments, making the outcome more transparent for intuitive grasp.

%
%xxxxxxxxxxxxxxxxxxxxxxxxxxxxxxxxxxxxxxxxxxxxxxxxxxxxxxxxxxxxxxxxxxxxxxxxxxxxxx
%
%
%%%%%%%%%%%%%%%%%%%%%%%%%%%%%%%%%%%%%%%%%%%%%%%%%%%%%%%%%%%%%%%%%%%%%%%%%%%%%%
\subsection{Non-interacting $XX$ limit ($\Delta=0$) }\label{XXsec}
%%%%%%%%%%%%%%%%%%%%%%%%%%%%%%%%%%%%%%%%%%%%%%%%%%%%%%%%%%%%%%%%%%%%%%%%%%%%%%
%
%
Most of the formulas of Sec.~\ref{D0Res} for free fermions can be brought to a closed form of standard mathematical functions.
The content of this subsection is implicitly present in the earlier work \cite{Chitov:2019}, but the $XX$ limit
was not specifically analyzed in that paper. The spectrum \eqref{Epm} becomes
\begin{equation}
\label{Epmxi}
 E_{\pm}(k)=h \pm \xi,~~ \xi \equiv  \sqrt{h_a^2+ t^2 \cos^2k+ \delta^2 \sin^2k}~.
\end{equation}

\begin{figure}[]
\centering{\includegraphics[width=7.0cm]{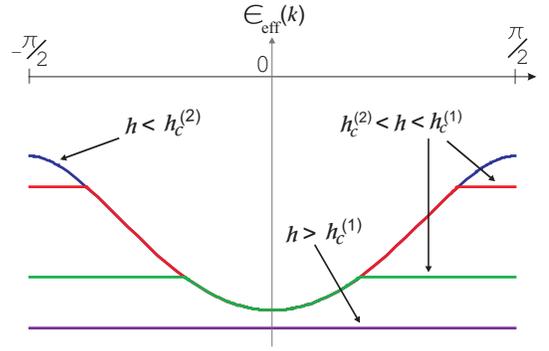}}
\caption{The effective single-particle spectrum $\varepsilon_{\text{eff}}(k)$ in three  phases at different values of the uniform field.  }
\label{FlatBand}
\end{figure}

To better understand results of this section, it is convenient to write the ground state energy per site
\begin{equation}
\label{E0}
f=  -\frac{1}{2 \pi}\int^{\pi/2}_{0} \big( |E_+| +|E_-| \big) dk
\end{equation}
as
\begin{equation}
\label{E0eff}
f=  \frac{1}{2 \pi}\int^{\pi/2}_{-\pi/2} \varepsilon_{\text{eff}}(k) dk.
\end{equation}
The effective spectrum $\varepsilon_{\text{eff}}(k)$ is shown in Fig.~\ref{FlatBand} for three phases.
From \eqref{Epmxi} and \eqref{E0} we find the $h$-independent effective spectrum $\varepsilon_{\text{eff}}(k)= -\xi$ in the
topological phase ($h<h_c^{\s (2)}$). In the IC gapless phase ($h_c^{\s (2)}<h<h_c^{\s (1)}$), the parabolic spectrum $\varepsilon_{\text{eff}}(k)= -\xi$ at $|k|<k_{\s F}$ with the Fermi momentum given by Eq.~\eqref{kIC},  becomes a flat band $\varepsilon_{\text{eff}}(k)=-h$ at $ k_{\s F}< |k|< \pi/2$. The Fermi sea shrinks with the growth of the field, as shown in Fig.~\ref{FlatBand}, and in the PM phase ($h>h_c^{\s (1)}$) the whole band is flat, $\varepsilon_{\text{eff}}(k)=-h$.

Analytically, we find:
\begin{equation}
\label{E0An}
 f=
\left\{
\begin{array}{lr}
-\frac12 h ,~ &h>h_c^{\s (1)} \\[0.3cm]
  -\frac{1}{\pi} \sqrt{t^2+h_a^2} \mathbf{E}(k_{\s F},\kappa^2)   -  \\ [0.15cm]
   -\frac12 h \big( 1- \frac{2}{\pi} k_{\s F} \big),~&h \in [h_c^{\s (2)}, h_c^{\s (1)}] \\[0.3cm]
 -\frac{1}{\pi} \sqrt{t^2+h_a^2} \mathbf{E}(\kappa^2),~ &h<h_c^{\s (2)}\\
\end{array}
\right.
\end{equation}
Here  $\mathbf{E}$  is the elliptic integral of the second kind, and
\begin{equation}
\label{kap}
  \kappa^2 \equiv \frac{t^2-\delta^2}{t^2+h_a^2}.
\end{equation}

The uniform magnetization derived from Eq.~\eqref{mz1}
\begin{equation}
\label{mzXX}
m_z=\frac12 +\frac{1}{\pi}\int^{\pi/2}_{0} \mathrm{sign} (E_-)  dk
\end{equation}
or obtained directly from differentiation of \eqref{E0An}, demonstrates two plateaux in the gapped phases,
connected by a continuous curve in between:
\begin{equation}
\label{mzAn}
 m_z=
\left\{
\begin{array}{lr}
1,~ &h>h_c^{\s (1)} \\ [0.2cm]
1- \frac{2}{\pi} k_{\s F}, ~&h \in [h_c^{\s (2)}, h_c^{\s (1)}] \\ [0.2cm]
0,~ &h<h_c^{\s (2)}\\
\end{array}
\right.
\end{equation}
The above results is in agreement with the arguments of Ref.~\cite{Affleck:1997}, generalizing the Lieb-Schultz-Mattis (LSM) theorem  \cite{LiebSM:1961} for non-zero field. According to another formulation of the LSM theorem in terms of fermions  \cite{Oshikawa:2000} (cf. Eq.~\eqref{MMa}), the plateaux of magnetization correspond to integer fermionic fillings per unit cell, and the filling can admit non-integer values only
in the gapless phase, leading to a smooth evolution of $m_z \in [0,1]$ at $h \in [h_c^{\s (2)}, h_c^{\s (1)}]$.

\begin{figure}[]
\centering{\includegraphics[width=7.5cm]{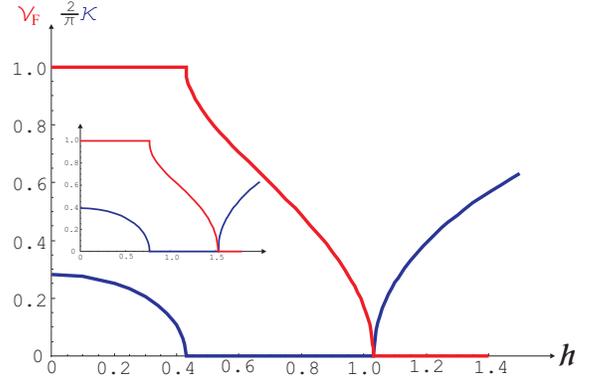}}
\caption{Real and imaginary parts of the complex wave vector $k_+$ giving the values of the fermionic filling per unit cell
$\nu_{\s F}=2/\pi k_{\s F}$ and inverse correlation length $\xi^{-1} \propto \kappa$ for the three phases. The main plot is done for non-interacting model with $h_a=0.25$ and $\delta=0.35$. The inset shows the same parameters for the interacting case with $\Delta=0.5$.  }
\label{ComplexKf}
\end{figure}

To unify and generalize the analysis of phases done in Sec.~\ref{D0Res}, and to directly relate it to the LSM theorem \cite{LiebSM:1961,Oshikawa:2000}, we
analytically continue the spectrum of the model onto the complex plane $z \in \mathbb{C}$ with $z = e^{ik}$ \cite{Franchini:2017}.
In the isotropic limit the eigenvalues $\lambda_\pm$  defined by Eq.~\eqref{lambdapm} become the eigenvalues of the Hamiltonian \eqref{Epmxi},
so the condition of the quantum criticality  \eqref{C4QCP} with Eq.~\eqref{C4Lpm} reads
\begin{equation}
\label{EpmZ}
  | E_+(z)|^2 | E_-(z)|^2=0~.
\end{equation}
Using $Q$ defined in Eq.~\eqref{kIC} and extended to $Q \in \mathbb{C}$, we find two roots of  \eqref{EpmZ}:
\begin{equation}
\label{Zpm}
z_\pm =e^{ik_\pm} =\Lambda_\pm,~\text{with}~ \Lambda_\pm \equiv iQ \pm \sqrt{1-Q^2}~,
\end{equation}
The roots  $\Lambda_\pm$ encode important information about three phases:\\
\textit{(1)} In the IC phase $Q \in \mathbb{R}$ and $0<Q<1$. The roots are complex conjugate $\Lambda_+=\Lambda_-^\ast$ and $|\Lambda_\pm|=1$.
The wave vectors $k_\pm \in \mathbb{R}$ and we can pick $k_+=k_{\s F}$ corresponding to the known solution \eqref{kIC}. The real wave vector
$k_{\s F}$ defines the period of oscillations of correlation functions (see Eq.~\eqref{DzzIC0} below) and controls the (IC) filling (Fermi level) of the parabolic
band $\nu_{\s F}=2/\pi k_{\s F}$, see Figs.~\ref{FlatBand} and \ref{ComplexKf}.\\
\textit{(2)} In the PM phase $Q=i |Q|$ is imaginary, and it leads to the imaginary $k_\pm=-i \ln \Lambda_\pm$, see Fig.~\ref{ComplexKf}:
\begin{eqnarray}
  \label{KfPM}
  k_{\s F} &=& \text{Re} k_\pm=0, \\
  \label{KapPM}
  \kappa  &=&  \text{Im}k_+=-\ln \big( \sqrt{1+|Q|^2}-|Q| \big)~.
\end{eqnarray}
The non-vanishing imaginary part of the complex root $k_+$ gives the inverse correlation length \cite{Franchini:2017}, and it is responsible for the exponential decay of correlation functions in gapped phases. In the vicinity of the PM transition $h\to h_c^{\s (1)}+0$: $|Q| \ll 1$, and
$\kappa \approx  |Q| \propto  (h- h_c^{\s (1)})^{1/2}$. The vanishing real part of the root $k_{\s F}=0$ means monotonous behavior of correlation functions without oscillations.

Note that probing the correlation length in the limit $\gamma \to 0$ is subtle in the PM (polarized) phase. In this case
$m_z=\mathcal{O}_z=1$ and $m_x=m_y=0$. Moreover,  the correlation functions are  featureless, i.e., $\langle \sigma^z_L \sigma^z_R \rangle=1$ and  $\langle \sigma^x_L \sigma^x_R \rangle= \langle \sigma^y_L \sigma^y_R \rangle=0$, $\forall L,R$. (These functions were first found by Barouch and McCoy
in \cite{McCoy:1971} for the case $h_a=\delta=0$.) The string correlation functions are found to behave in a similar way, i.e., $\mathcal{D}_{zz}=1$ and $\mathcal{D}_{xx}=\mathcal{D}_{yy}=0$  at $\gamma=0$.  As one can find in Table 1.2 of the book by Franchini  \cite{Franchini:2017} at $h>1$ and $\gamma \neq 0$ ($h_a=\delta=0$), the spin correlation function
\begin{equation*}
%\label{xx}
  \langle \sigma^x_1 \sigma^x_n \rangle \simeq   X_D \frac{\Lambda_+^{-n}}{\sqrt{n}}~~\mathrm{at} ~~ n \gg 1~,
\end{equation*}
where $\Lambda_+$ defined as in Eq.\eqref{Zpm}, determines the correlation length and agrees with the $\gamma \to 0$
result \eqref{KapPM}. Vanishing of the correlation function in the above equation is due to prefactor $X_D \to 0$ as $\gamma \to 0$ \cite{Franchini:2017}.
No analytical results are available for the spin or string correlation functions in the general case $h_a \neq 0$ and
$\delta \neq 0$, but we infer from numerical calculations that those functions have: \textit{(i)} rapid
decrease with $n$; \textit{(ii)} vanishing amplitudes; \textit{(iii)} finite gap and thus meaningful definition of the correlation length in the limit $\gamma \to 0$, similar to the equation above. Another way to probe the finite gap (inverse correlation length) at $\gamma=0$, is to consider non-zero temperature, when one expects temperature corrections to the correlation and/or response functions   $\propto \exp(-\mathcal{O}(1) \kappa /T)$.
\\
\textit{(3)} For the $\mathcal{O}_z(\pi/2)$-phase it is convenient to use $V \equiv \sqrt{1-Q^2}$. In this phase  $V=i |V|$ is imaginary, leading to
\begin{eqnarray}
  \label{KfOz}
  k_{\s F} &=& \text{Re} k_+=\frac{\pi}{2}, \\
  \label{KapOz}
  \kappa  &=&  -\text{Im}k_+=\ln \big( \sqrt{1+|V|^2}+|V| \big)~.
\end{eqnarray}
The real part of the root $k_{\s F}=\pi/2$ corresponds to the constant filling $\nu_{\s F}=1$ in this phase and $\pi/2$-oscillations of the
string correlation function \eqref{DzplAn}. Near transition point  $h\to h_c^{\s (2)}-0$: $|V| \ll 1$, and
$\kappa \approx  |V| \propto  (h_c^{\s (2)}-h)^{1/2}$, in agreement with the expected gap closing.

The above results for real and imaginary parts of the complex roots $k_\pm $ are depicted in Fig.~\ref{ComplexKf}. In agreement with general arguments \cite{Oshikawa:2000} and with Fig.~\ref{ComplexKf}, the magnetization in all three phases can be related to the filling as  $m_z=1-\nu_{\s F}$.

Qualitatively, the integer-valued fermionic  fillings  connected by a continuous curve through the gapless phase of Fig.~\ref{ComplexKf},
are due to the flat piece of the effective spectrum shown in Fig.~\ref{FlatBand}.  An interesting topological transition known as fermionic condensation \cite{KhodelShaginyan:1990} is signalled by appearance of a continuous real filling smoothly connecting between two integer values (1,0) of the step function predicted by the Landau Fermi-liquid theory. One needs a flat band piece of the single-particle spectrum for such non-integer filling to occur.
The flat band and fermionic condensation \cite{Volovik:2019} can model linear-$T$ resistivity in the so-called Planckian metal \cite{Shaginyan:2019,Sachdev:2019}. These analogies are worth exploring further.

Two complex roots $\Lambda_\pm$ are also the eigenvalues of the transfer matrix which generates the wave function of the zero-energy edge Majorana fermion
\cite{Chitov:2018}. Our findings predict that the localized Majorana edge state in the $\mathcal{O}_z(\pi/2)$-phase has the wave function with the inverse penetration depth $\propto \kappa$. The exponential decay of the wave function into the bulk is  modulated by $\pi/2$-oscillations. In the IC phase the edge state gets delocalized, since $\kappa=0$.

The staggered magnetization found from Eq.~\eqref{mza1} as
\begin{equation}
\label{mzaXX}
m_z^a= \frac{h_a}{\pi}\int^{\pi/2}_{0} \frac{dk}{\xi} \big( 1- \mathrm{sign} (E_-) \big)~.
\end{equation}
leads to
\begin{equation}
\label{mzaAn}
 m_z^a=
\left\{
\begin{array}{lr}
0,~ &h>h_c^{\s (1)} \\[0.2cm]
 \frac{2 h_a}{\pi \sqrt{t^2+h_a^2}} \mathbf{F}(k_{\s F},\kappa^2) , ~&h \in [h_c^{\s (2)}, h_c^{\s (1)}] \\[0.2cm]
 \frac{2 h_a}{\pi \sqrt{t^2+h_a^2}} \mathbf{K}(\kappa^2),~ &h<h_c^{\s (2)}\\
\end{array}
\right.
\end{equation}
$\mathbf{K}$ and $\mathbf{F}$ are, respectively, the complete and incomplete elliptic integrals of the first kind.

The bond average and dimerization susceptibility \eqref{teta} can be also found in a closed form via elliptical functions \cite{Chitov:2004PRB}.
Indeed,
\begin{equation}
\label{KXX}
\mathcal{K} = \frac{1}{2 \pi}\int^{\pi/2}_{0} dk \frac{\cos^2 k}{\xi} \big( 1- \mathrm{sign} (E_-) \big)~
\end{equation}
yields
\begin{widetext}
\begin{equation}
\label{KAn}
 \mathcal{K}=
\left\{
\begin{array}{lr}
0,~ &h>h_c^{\s (1)} \\[0.2cm]
 \frac{\sqrt{t^2+h_a^2}}{\pi(t^2-\delta^2)} \big[ \mathbf{E}(k_{\s F},\kappa^2)-(1-\kappa^2) \mathbf{F}(k_{\s F},\kappa^2) \big] ,
  ~&h \in [h_c^{\s (2)}, h_c^{\s (1)}] \\[0.2cm]
 \frac{\sqrt{t^2+h_a^2}}{\pi(t^2-\delta^2)} \big[ \mathbf{E}(\kappa^2)-(1-\kappa^2) \mathbf{K}(\kappa^2) \big] ,
  ~&h<h_c^{\s (2)}\\
\end{array}
\right.
\end{equation}
\end{widetext}
The dimerization susceptibility
\begin{equation}
\label{etaXX}
\eta= \frac{1}{2 \pi}\int^{\pi/2}_{0} dk \frac{\sin^2 k}{\xi} \big( 1- \mathrm{sign} (E_-) \big)~,
\end{equation}
is found as
\begin{equation}
\label{etaAn}
 \eta =
\left\{
\begin{array}{lr}
0,~ &h>h_c^{\s (1)} \\[0.2cm]
 \frac{\sqrt{t^2+h_a^2}}{\pi(t^2-\delta^2)} \big[ \mathbf{F}(k_{\s F},\kappa^2)- \mathbf{E}(k_{\s F},\kappa^2) \big] ,
  ~&h \in [h_c^{\s (2)}, h_c^{\s (1)}] \\[0.2cm]
 \frac{\sqrt{t^2+h_a^2}}{\pi(t^2-\delta^2)} \big[ \mathbf{K}(\kappa^2)- \mathbf{E}(\kappa^2) \big] ,
  ~&h<h_c^{\s (2)}\\
\end{array}
\right.
\end{equation}
In the isotropic limit two anomalous parameters  \eqref{Peta} breaking the particle number conservation, $P=\eta_{\s P}=0$ due to $U(1)$ symmetry.

Some additional progress in analytical evaluation of the string correlation function
\eqref{Dzz} can be made for $XX$ chain.
The Majorana correlation function \eqref{abS} gets simplified. Introducing
\begin{equation}
 \label{Gpm}
  G^\pm (k) \equiv  G_{11}(k) \pm G_{12}(k)
\end{equation}
and
\begin{equation}
 \label{gpm}
  g^\pm  \equiv  (t \cos k \pm h_a \pm i \delta \sin k)/\xi~,
\end{equation}
we find
\begin{equation}
\label{GpmXX}
G^\pm (k) = \frac12 \big(1+ \mathrm{sign} (E_-) \big) +\frac12   \big(1- \mathrm{sign} (E_-) \big)g^\pm
\end{equation}
The above equation yields for the gapped phases
\begin{equation}
\label{GpmAn12}
 G^\pm (k)=
\left\{
\begin{array}{lr}
1,~ &h>h_c^{\s (1)} \\[0.2cm]
g^\pm, ~&h<h_c^{\s (2)} \\
\end{array}
\right.
\end{equation}
and for the gapless IC phase at $h_c^{\s (2)} <h<h_c^{\s (1)}$:
\begin{equation}
\label{GpmAn3}
 G^\pm (k)=
\left\{
\begin{array}{lr}
1,~ &k>k_{\s F} \\[0.2cm]
g^\pm, ~&k<k_{\s F}  \\
\end{array}
\right.
\end{equation}

At $h>h_c^{\s (1)}$ the block Toeplitz matrix for evaluation of  $\mathfrak{D}_{zz}$
(cf. Ref.~\onlinecite{Chitov:2019} its explicit form) becomes just a unit matrix for any choice
of $L$ and $R$ in \eqref{Dzz}. So we find the exact result for the SOP:
\begin{equation}
\label{Oz1}
  \mathcal{O}_z=1, ~~h>h_c^{\s (1)}~.
\end{equation}
The above result for the average of strings of $\sigma^z$ operators \eqref{Oz} is in sync with the existence of plateau of magnetization
$m_z =\langle \sigma^z \rangle=1$.

At $h<h_c^{\s (2)}$ the SOPs $\mathcal{O}_{z,i}$ form step-like  parabolic lines along $h$, similar to  Eqs.~(\ref{mzaAn},\ref{KAn},\ref{etaAn})
\cite{Watanabe:2018}.
The values of $\mathcal{O}_{z,i}$ are available via numerical calculations only. However, in case $h_a=0$ the result (\ref{Oz1dg}) can be used
to find SOP inside the circle along the line $\gamma=0$:
\begin{equation}
\label{Oz1g0}
 h_a=\gamma=0,~|h|<h_c^{\s (2)}:~~ \mathcal{O}_{z,1}^2=2 \frac{\delta^{1/2}}{1+\delta}.
\end{equation}
The model at $\gamma=0$ with additional $U(1)$ symmetry belongs to a separate universality class with the central charge $c=1$ \cite{Franchini:2017}.
From \eqref{Oz1g0} we infer the index of the order parameter $\beta = 1/4$ in the vicinity of the critical point $\delta=0$.\footnotemark[1]
Unfortunately, no progress is made at this point in analytical evaluation of SOPs beyond two special cases (\ref{Oz1dg},\ref{Oz1g0}).

Two plateaux of $m_z$ have a certain analogy with quantized Hall conductance, proportional to the topological Chern number \cite{Niu:1985}.
In the isotropic limit the eigenvalues $\lambda_\pm$  defined by Eq.~\eqref{lambdapm} become the eigenvalues of the Hamiltonian \eqref{Epmxi}.
In such case the winding number  \eqref{NwAn} and magnetization \eqref{mzXX} are simply related in the gapped phases:
\begin{equation}
\label{mzNw}
  m_z=1-N_w
\end{equation}

The IC gapless phase does not have long-range string order, since all three $\mathcal{O}_{z,i}=0$ on the right hand side of \eqref{Ozeo}.
(In the limit $\gamma \to 0$, parameters $\mathcal{O}_{z,i}$ vanish abruptly as $h \to h_c^{\s (2)} +0$ and $h \to h_c^{\s (1)}-0$,
see Fig.~\ref{OPsall}(a) for vizualization).
However the gapless phase is \textit{algebraically ordered}, demonstrating power-law decaying string-string correlation function with the IC oscillations:
\begin{equation}
\label{DzzIC0}
  \mathfrak{D}_{zz}(1,n) = \frac{\mathcal{A}}{\sqrt{n}} \cos(k_{\s F} n)~.
\end{equation}
In the above formula the coefficient $\mathcal{A}$ is non-universal, while the critical index of the correlation function $\eta =1/2$. The latter along with other two indices $\nu =1$ and $\beta =1/4$ satisfy all scaling relations. We found a perfect agreement between Eq.~(\ref{DzzIC0}) and direct
numerical calculation of the string correlation function.  For a particular choice of parameters yielding $k_{\s F}=\pi/6$, the results are shown in Fig.~\ref{SOPOscil}(a) with $\mathcal{A}=1/\pi^{1/8} \approx 0.87$.

%
%
%%%%%%%%%%%%%%%%%%%%%%%%%%%%%%%%%%%%%%%%%%%%%%%%%%%%%%%%%%%%%%%%%%%%%%%%%%%%%%
\subsection{Interacting $XXZ$ limit ($\Delta \neq 0$)}\label{XX1}
%%%%%%%%%%%%%%%%%%%%%%%%%%%%%%%%%%%%%%%%%%%%%%%%%%%%%%%%%%%%%%%%%%%%%%%%%%%%%%
%
%

%
%
%%%%%%%%%%%%%%%%%%%%%%%%%%%%%%%%%%%%%%%%%%%%%%%%%%%%%%%%%%%%%%%%%%%%%%%%%%%%%%
\subsubsection{Plateaux, parabolic lines, string order, and oscillations}\label{XXRen}
%%%%%%%%%%%%%%%%%%%%%%%%%%%%%%%%%%%%%%%%%%%%%%%%%%%%%%%%%%%%%%%%%%%%%%%%%%%%%%
%
%

To deal with the regime of weak interactions $\Delta \alt 1$ we need to replace the bare parameters in the equations of the previous subsection
by the renormalized quantities. Having almost all results expressed via standard functions does not rescind the task of extensive numerical calculations, since critical fields  $h_c^{\s (1,2)}$ and remormalized couplings must be found from self-consistent mean-field equations for each
point in the four-dimensional space of bare parameters. (Two parameters are eliminated from our analysis, since we found $\gamma_{\s R}= \gamma_{a \s R}=0$ in the isotropic limit.)

As expected \cite{Affleck:1997}, the uniform magnetization in the interacting model has two plateaux, as in Eq.~\eqref{mzAn}. It is shown in Fig.~\ref{Mzall}. The interaction renormalizes numerical values of critical fields  $h_c^{\s (1,2)}$ and the form of the curve $m_z$ in the IC gapless phase, but not the universal plateau values $m_z=0,1$ in two gapped phases. Qualitatively, it means that the interaction does not change the integer-valued filling per unit cell (1 or 0), in agreement with Ref.~\cite{Oshikawa:2000},  as one can see in Fig.~\ref{ComplexKf}.
Thus, the present mean-field theory respects the LSM theorem.

The analytical results of the previous subsection allow us to understand how this interesting feature makes it way through the equations.
The renormalized effective spectrum has the same form, as shown in Fig.~\ref{FlatBand}.
The bond average $\mathcal{K}$ and dimerization susceptibility $\eta$ (cf. \eqref{KAn} and \eqref{etaAn})
in the $\mathcal{O}_z(\pi/2)$-phase  ($h< h_c^{\s (2)}$) are $h$-independent functions of other couplings, see Fig.~\ref{MFPars}(a).
In mathematical terms, these functions form \textit{parabolic lines} of zero curvature in the parametric space. Both quantities  $\mathcal{K}$ and $\eta$
vanish in the PM phase  ($h> h_c^{\s (1)}$). The staggered magnetization $m_z^a$, cf. \eqref{mzAn} and Fig.~\ref{Mzall}, demonstrates similar behavior.
The ground state energy $f$ in the $\mathcal{O}_z(\pi/2)$-phase is given by the third expression in \eqref{E0An} plus the constant term
$\Delta \mathcal{C}$  where  $\mathcal{C}$, determined from Eq.~\eqref{Const}, is
\begin{equation}
\label{Ctop}
\mathcal{C}= \mathcal{K}^2 + \frac14 (m_z^a)^2 + \delta^2 \big( \eta^2  +2  \mathcal{K} \eta \big)~.
\end{equation}
The above term and $h$-independent effective band $\varepsilon_{\text{eff}}$  lead to $f$ as an $h$-independent parabolic line at $h< h_c^{\s (2)}$
\cite{[{The results for the $h$-independent quantities in the interacting and non-interacting cases, are in agreement with the results of Watanabe:
expectation values of several observables do not depend on magnetic flux in the gapped phases of the model with
U(1) symmetry, see }] Watanabe:2018},
and, consequently, $m_z=0$.
For the PM phase  with  a totally flat band we find
\begin{equation}
\label{fxx}
  f=-\frac14 \Delta -\frac12 h~,
\end{equation}
where Eq.~\eqref{hR} with $h_{\s R}=h-\Delta$ is used, leading to $m_z=1$
\cite{[{The first exact result on magnetization plateau in the $XXX$ model with uniform field was reported in }] Griffiths:1964}.
So, the plateau $m_z=1$ is due to: (1) flat band which leads to linear dependence of the ground-state energy on the field; (2) the fact that interaction does not renormalize the slope (-$1/2$) of this straight line. The relation between the magnetization and the winding number \eqref{mzNw} holds for the interacting case.

The exact result \eqref{Oz1} holds for the plateau of the SOP $\mathcal{O}_z$ in PM phase of the interacting model, along with the step-like $h$-independent behavior of three parameters $\mathcal{O}_{z,i}$ of the oscillating string long-ranged order, see Fig.~\ref{OPsall}(a) and Fig.~\ref{SOPDetails}. Since the interaction does not change the wave vector of the oscillating string order $k_{\s F}=\pi/2$, cf.
Fig.~\ref{ComplexKf}, one can select a convenient single correlation function  \eqref{DzplAn} and to use the SOP $\mathcal{O}_{z,+}$.

It is worth stressing qualitative similarities and distinctions in behaviors of the average quantities entering our equations in two gapped phases: While in the gapped topologically trivial PM phase ($h>h_c^{\s (1)}$) all quantities $m_z,m_z^a,\mathcal{O}_{z,i},\mathcal{K},\eta$ are equal to 1 or 0, i.e., they form true plateaux, in the gapped topological phase ($h<h_c^{\s (2)}$) only the uniform magnetization demonstrates a true (trivial) plateau $m_z=0$. The other quantities are $h$-independent functions of other couplings  (parabolic lines) \cite{Watanabe:2018}.
Except for the SOPs, all other quantities are having their values in the gapped phases continuously connected  across the IC gapless phase with cusps at two critical points $h_c^{\s (1)}$ and  $h_c^{\s (2)}$.

\begin{figure}[]
\centering{\includegraphics[width=7.5cm]{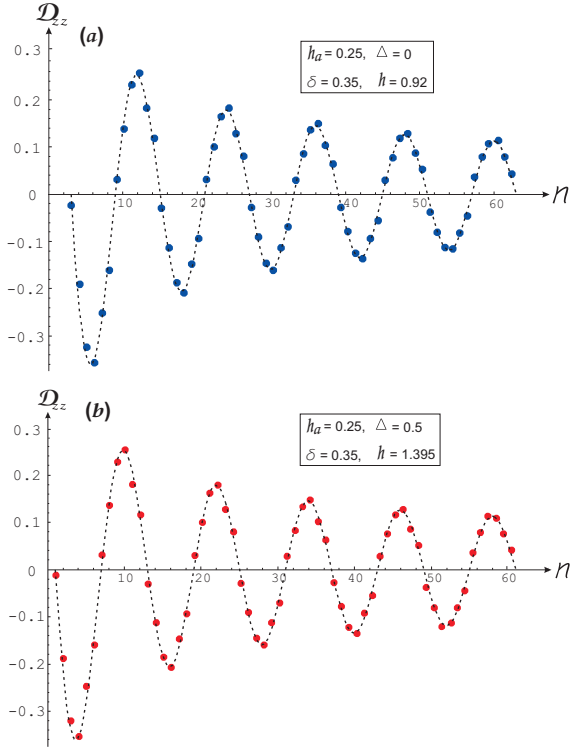}}
\caption{Power-law decaying oscillations of the string correlation function $\mathfrak{D}_{zz}(1,n)$ in the gapless IC phase.
Panel (a) shows direct numerical results from the Toeplitz determinant (blue dots) for the non-interacting case and the
plot of  Eq.~\eqref{DzzIC0} (dashed line).
Panel (b) shows direct numerical results for the interacting case when the Toeplitz determinant is calculated with renormalized parameters
found from the mean-field equations (red dots) and the plot of  Eq.~\eqref{DzzIC} (dashed line)
In both cases $\mathcal{A}=1/\pi^{1/8}\approx 0.87$ and $k_{\s F} = \pi/6$.}
\label{SOPOscil}
\end{figure}

The IC gapless phase is the Luttinger liquid (LL) of the JW fermions \cite{Luther:1975}
\footnote{The mean-field approximation cannot account for such reconstruction of the fermionic ground state due to interactions. Instead of LL, the mean field predicts free fermions, albeit with renormalized parameters. In a moderately minimalist sense we accept this approximation as adequate, since the exact approach and the mean field both predict a gapless fermionic state.}
The long-range string order of the gapped topological phase ($h<h_c^{\s (2)}$) is taken over by the algebraic order of the power-law decaying string correlations at $h_c^{\s (2)} <h< h_c^{\s (1)}$. We have verified numerically oscillating behavior of \eqref{Dzz}. It is in agreement with predictions \eqref{DzzIC0}.\footnotemark[1] In the mean-field approximation the only effect of interactions on the correlation function is renormalization of $k_{\s F}$ and a phase shift. For a comparison with the non-interacting case \eqref{DzzIC0} presented in Fig.~\ref{SOPOscil}(a), we chose the model parameters to make the renormalized $k_{\s F}=\pi/6$ again. The direct numerical calculations are in excellent agreement with the analytical expression
\begin{equation}
\label{DzzIC}
  \mathfrak{D}_{zz}(1,n) = \frac{\mathcal{A}}{\sqrt{n+2}} \cos \Big( \frac{\pi}{6} (n+2) \Big)~,
\end{equation}
as one can see from Fig.~\ref{SOPOscil}(b). The gapless IC (LL) phase is a counterpart of the floating phase occurring via a BKT thermal phase transition in frustrated 2D Ising models  \cite{VillainBak:1981,*Bak:1982,Chitov:2005,*Chitov:2013}.
%
%

%
%
%%%%%%%%%%%%%%%%%%%%%%%%%%%%%%%%%%%%%%%%%%%%%%%%%%%%%%%%%%%%%%%%%%%%%%%%%%%%%%
\subsubsection{Interaction-driven transition}\label{MIT}
%%%%%%%%%%%%%%%%%%%%%%%%%%%%%%%%%%%%%%%%%%%%%%%%%%%%%%%%%%%%%%%%%%%%%%%%%%%%%%
%
%
The main goals of this subsection is to establish restrictions of the proposed mean-field theory
coming from the strength of interactions $\Delta$, and to relate the predicted $\mathcal{O}_z(\pi/2)$-phase
to the antiferromagnetic phase known from exact solution. The model of this study is antiferromagnetic, so
$\Delta>0$. Since we are interested to probe effects of the interaction, we turn off other relevant couplings and set
$h_a=\delta=0$.

\begin{figure}[]
\centering{\includegraphics[width=6.5cm]{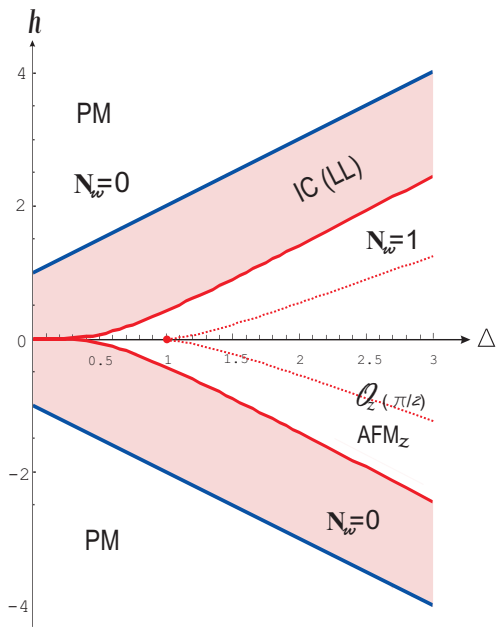}}
\caption{Phase diagram of the model with zero bare staggered field and dimerization in ($h,\Delta$)-plane. The exact result \cite{Takahashi:1999,Franchini:2017} for the phase boundary between the phase $\text{AFM}_z$ with spontaneous staggered magnetization and gapless IC (LL) phase, ending at the critical point (0,1), is indicated with red dotted line. The bold red line is the mean-field prediction for this boundary. The mean-field approximation agrees with the exact result for the IC-PM phase boundary, shown in bold blue. The topological winding numbers $N_w$ are also shown for each phase.}
\label{DeltaDiag}
\end{figure}

We analyse the model on the $(h,\Delta)$-plane shown in Fig~\ref{DeltaDiag}.
As known from exact results \cite{Takahashi:1999,Franchini:2017}, the chain without external fields generates
spontaneous antiferromagnetism  ($\text{AFM}_z$) in the axial direction ($m_z^a \neq 0$) at the critical value $\Delta=1$. At the non-interacting point $\Delta=0$ the model is in the IC phase at $0<h<1$, as we infer from Fig.~\ref{PDiag} along the line $\gamma=0$ (the $\mathcal{O}_z(\pi/2)$ circle is absent, since $\mathcal{R}(h_a=\delta=0)=0$). At $h=1$ the non-interacting model enters the familiar PM phase.  The IC-PM  phase boundary
$h_c^{\s (1)}=1+\Delta$  is a special case of the exact result \eqref{hc1XXZ}.
At $\Delta>1$ the $\text{AFM}_z$-phase resides inside  the V-shaped wedge on the $(h,\Delta)$-plane, and
at a certain critical field $h=h_c^{\s (2)}$ the $XXZ$ chain undergoes a phase transition into the IC (LL) phase. This phase boundary,
known exactly from the Bethe ansatz, is  schematically shown in Fig~\ref{DeltaDiag}.

The interaction is a marginal perturbation of the free fermionic Hamiltonian. One can check from the mean-field equations that along with
the trivial solution $m_z^a$, corresponding to the gapless IC state, consistent with the exact results at $\Delta<1$, those equations admit
a non-trivial solution $m_z^a \neq 0$ corresponding to the spontaneously generated antiferromagnetism. The order parameter of this phase
is the spontaneous staggered magnetization and it can be found analytically in the regime of weak interaction:
\begin{equation}
\label{mzaSSB}
   m_z^a \approx \frac{2}{\Delta} \exp \Big(-\frac{\pi}{2 \Delta} \Big)~, ~~\Delta \lesssim 1~.
\end{equation}
At large $\Delta \gg 1$ the order parameter saturates towards  $m_z^a \sim 1$.
A non-trivial  $m_z^a$  generates via Eq.\eqref{haR} the spontaneous staggered field $h_{a \s R}=\Delta m_z^a$.

We have checked that at the critical value $\Delta=1$ and $h=0$, the mean field
predicts the ground-state energy of the  $\text{AFM}_z$-phase $f_{\s \text{AFM}_z}=-0.4323$, while for the gapless IC-phase  $f_{\s \text{IC}}=-0.4196$ with the relative gain of the $\text{AFM}_z$-phase about $3 \%$. At $\Delta=1/2$ and $h=0$ the parameters are $f_{\s \text{AFM}_z}=-0.3694$ and
$f_{\s \text{IC}}=-0.3690$, with the relative gain  $\sim 0.1 \%$. At smaller $\Delta$ the gain is even smaller, and the two states are virtually degenerate.
However an unbiased minimization predicts at $h=0$ the winning antiferromagnetism all the way to $\Delta=0$, albeit exponentially weak \eqref{mzaSSB}.
At $\Delta \lesssim 1$ the mean field predicts the $\text{AFM}_z$-IC phase boundary
\begin{equation}
\label{hc2AFM}
  h_c^{\s (2)} \sim h_{a \s R} \approx  2 \exp \Big(-\frac{\pi}{2 \Delta} \Big)~,
\end{equation}
while at $\Delta \gg 1$ the critical field $h_c^{\s (2)}$ crosses over towards
\begin{equation}
\label{hc2big}
  h_c^{\s (2)} \propto \Delta~.
\end{equation}
The result of numerical mean-field calculations for $h_c^{\s (2)}$ is shown in Fig.~\ref{DeltaDiag}.

The $\text{AFM}_z$-IC phase transition is of the first order in the mean-field theory, since $m_z^a$ undergoes a jump from zero in the gapless IC
phase to a finite value and stays constant for a given $\Delta$ in $\text{AFM}_z$-phase ($h<h_c^{\s (2)}$). (Note that $\Delta=$~const lines on the $(h,\Delta)$-plane are parabolic lines of $h$-independent parameters, like $f,m_z^a,h_{a \s R}$, etc, as explained above.)
The spontaneous antiferromagnetism with its primary order parameter $m_z^a$ coexists with the four-periodic string order defined in previous sections.
The SOPs are induced by the staggered field $h_{a \s R} \propto  m_z^a$ and are the secondary. The order is
of the type  \eqref{delta0} with patterns shown in panels (2a) and (2b) in Fig.~\ref{SOPDetailsSplit}. The string order can be also combined into a single
pattern as in Fig.~\ref{SOPDetailsSplit} (1) with the help of the correlation function \eqref{DzplAn}.
The algebraically ordered gapless IC(LL) phase in Fig.~\ref{DeltaDiag} is characterized by the power-law decaying correlation functions,
similar to the one shown in Fig.~\ref{SOPOscil}.

So the phase with the interaction-induced antiferromagnetism $\text{AFM}_z$ on the phase diagram in Fig.~\ref{DeltaDiag}, \textit{per se} is just a
special case of the $\mathcal{O}_z(\pi/2)$ phase shown in Fig.~\ref{PDiag}. That is why the second label  $\mathcal{O}_z(\pi/2)$
for the magneic phase is added in Fig.~\ref{DeltaDiag}. However, the transition into the gapless IC phase is quite different in two cases, revealing
important distinctions between the two phases. The axial symmetry broken in the interaction-generated $\text{AFM}_z$-phase which possesses  a sublattice magnetization and doubling of a unit cell, is restored via the first order transition into the IC (LL) phase. The latter, shown in  Fig.~\ref{DeltaDiag}, has both the staggered field and magnetization zero, $m_z^a= h_{a \s R}=0$. For the case of the field-generated  $\mathcal{O}_z(\pi/2)$-IC transition
shown in Fig.~\ref{PDiag}, \textit{no symmetry breaking related to sublattice (staggered) magnetization occurs, and the field-induced $m_z^a$ is not the order parameter.} It is continuous across transition and has a cusp only, as one can see in Fig.~\ref{Mzall}; the gaplessness of the IC phase is a result of subtle interplay of several relevant couplings.  In both cases of the $\text{AFM}_z$ or $\mathcal{O}_z(\pi/2)$ phases,  the uniform magnetization $m_z$ is zero, as shown in Fig.~\ref{Mzall} ($\gamma=0$) at $h<h_c^{\s (2)}$.

The last comment is in order here to address the validity of the proposed mean-field approach. More exactly: how the mean-field prediction of the spurious spontaneous antiferromagnetism with $h_a=\delta=0$  in the range $\Delta <1$ can undermine our predictions for the phase diagram in
Fig.~\ref{PDiag}? The answer is two-fold: in the absence of relevant terms $\propto h_a$ or $\propto \delta$ and $h=0$, the mean-field instability in the region $\Delta<1$ signals the need of that approximation to be replaced by more sophisticated techniques.  In the case when one or more of the mentioned parameters are non-zero, the (exponentially weak) interaction-generated terms do not drive spontaneous magnetization, but rather result in innocuous renormalizations of model's parameters. As an example tested by direct simulations, we can mention our earlier work on coupled dimerized $XXX$-chains ($\Delta=1$)  \cite{Chitov:2008,*Chitov:2011PRB,Chitov:2017JSM} where two relevant parameters -- dimerization and inter-chain coupling -- are present. The mean-field predictions are shown to be very accurate quantitatively, no spurious phases, in agreement with DMRG or exact diagonalization results, even on the lines of quantum criticality where a mutual cancellation of relevant terms occurs.

%xxxxxxxxxxxxxxxxxxxxxxxxxxxxxxxxxxxxxxxxxxxxxxxxxxxxxxxxxxxxxxxxxxxxxxxxxxxxxx
%
\section{Conclusion}\label{Concl}
%
%xxxxxxxxxxxxxxxxxxxxxxxxxxxxxxxxxxxxxxxxxxxxxxxxxxxxxxxxxxxxxxxxxxxxxxxxxxxxxx
%

The phase diagram and the order parameters of the $XYZ$ spin-$1/2$ chain with alternation of the exchange and anisotropy couplings  in the presence
of uniform and staggered magnetic fields are analyzed. In the fermionic representation the model is the interacting Kitaev-Majorana chain  with
hopping, superconducting pairing, and chemical potential modulated. The model is treated within the Landau mean-field framework, where the role of the Ginzburg-Landau potential is played by the effective quadratic fermionic Hamiltonian, derived from the Hartee-Fock (HF) approximation of the interacting
fermionic Hamiltonian of the model. The effective HF Hamilonian is expressed in terms of the renormalized couplings, ``dressed" by interactions, which are determined from minimization of the thermodynamic potential. In the non-interacting limit $\Delta=0$ the HF Hamiltonian recovers the exact one of the free JW fermions, and the renormalized couplings become the bare microscopic parameters of the model.

In this paper we have worked out all the steps of the framework to deal with an interacting  problem involving local and nonlocal orders within the same (extended) Landau formalism. The main progress with respect to the earlier related work \cite{Chitov:2017JSM,Chitov:2018,Chitov:2019}, is to present a solution for a physically interesting non-integrable model, to connect the tools available for the exactly-solvable quadratic fermionic Hamiltonians with the standard methods of the mean-field approximation.

The steps of analysis are as follows: \\
Since the effective Hamiltonian is quadratic, its eigenvalues can be found analytically. All possible phases of the model and conditions for the phase boundaries are found from zeros of the spectrum. In case of competing orders, the stable phase is determined by the energy minimum. More physically relevant information is available if analysis of zeros of the spectrum  is extended on the complex plane of wave numbers, however it is not always technically straightforward. In this study such analysis was done for the axial symmetric limit of the model. On the phase diagram of the model three possible local order parameters (components of the magnetization) and the nonlocal string order parameter are identified in general case. The local and nonlocal order parameters are expressed via the string correlation functions of Majorana fermions. The latter are evaluated as asymptotes of the determinants of the block Toeplitz matrices. For the effective quadratic Hamiltonian with six renormalized couplings, two unitary matrices of the Bogoliubov transformation were found. These matrices allow to derive an analytic expression for the correlation function of two Majorana fermions, which defines elements of those block Toeplitz matrices. These exact methods are combined
with the self-consistent approximation. The latter is a component of the mean-field theory (along with the decoupling and approximation of the Hamiltonian), which uses the minimization of the thermodynamic potential to determine the renormalized couplings (mean-field parameters) of the effective Hamiltonian.

The main result of the above formalism combining the exact and the mean-field methods, is the phase diagram of the model found numerically and shown in Fig.~\ref{PDiag}, and its local and string order parameters. The representative numerical results for the latter are plotted in Fig.~\ref{OPsall}. The predictions for conventional (local) orders agree with the earlier results \cite{Takahashi:1999,Franchini:2017,denNijs:1981,Alcaraz:1995,Okamoto:1996,Yamamoto:2000,Dmitriev:2002JETP,Dmitriev:2002PRB} available only for some special choices of parameters of the model we study. We found the topological phase on the diagram with oscillating string order with a period of four lattice spacings which was not reported before for this model. A detailed analysis of patterns of the string order is given.  In addition we have calculated the winding numbers $N_w$ for all phases. The phase with the oscillating topological SOP is the only one with non-trivial $N_w=1$. In particular, we have shown that the topological winding number cannot change without crossing gapless phase boundary. The present results agree with the recent results for the $XY$ chain \cite{Chitov:2019}, which is the non-interaction limit of the current model.

The $U(1)$-symmetric $XXZ$ limit of the model was given a special consideration.  It was demonstrated that the present approach respects the LSM theorem and its implications. In particular, plateaux and $h$-independent parabolic lines were revealed in various physical quantities, most notable, in the uniform axial magnetization, in accordance with general arguments \cite{Affleck:1997,Oshikawa:2000,Watanabe:2018}. The appearance of the integer-valued and IC fermionic fillings, responsible for qualitatively different behavior of the physical parameters in the gapped and gapless phases,
can be qualitatively related to the presence of flat band in the effective single particle spectrum. Turning on the anomalous $U(1)$-symmetry breaking coupling $\gamma \neq 0$, rounds the flat band and smears plateaux of magnetization and other step-like parameters.  The IC (LL) gapless phase with the algebraic order of power-law decaying correlations, is unstable versus any $\gamma \neq 0$, transforming into gapped phases with spontaneous planar magnetization ($m_{x,y}$, depending on the sign of $\gamma$). The topological order, which we associate with the oscillating SOP, evolves continuously (albeit not smoothly) through the $\gamma=0$ line inside the circle on the ($h,\gamma$)-plane, without gap closing, vanishing order parameter, or changing topological winding number. Similarly, nothing particular happens in the PM phase  $h>h_c^{\s (1)}$ along $\gamma=0$ line.

So, this line is a gapless line of quantum criticality only at $h_c^{\s (2)}<h<h_c^{\s (1)}$ separating gapped AFM phases with planar spontaneous magnetizations $m_{x,y}$. In the topological phase $\mathcal{O}_z(\pi/2)$ inside the circle, the SOP signalling discrete $\mathbb{Z}_2 \otimes \mathbb{Z}_2$ symmetry breaking, demonstrates four lattice spacing periodicity throughout. The line $\gamma=0$ inside this phase corresponds to additional $U(1)$-symmetry which brings about conserving quantities, but no transition changing the nature of the order in the  $\mathcal{O}_z(\pi/2)$-phase, is identified at $\gamma=0$.

The mean-field results of the present study lay a very good intuitively clear framework for further more technically sophisticated work.  Most importantly, direct numerics, like DMRG and/or exact diagonalization, plus heavier analytical guns, like RG and bosonization, are needed to check beyond the mean field the robustness of the predicted phase boundaries and stability of the phases in the sensitive parametric range, along with winding numbers and zero-energy Majorana edge states, with respect to the interaction-driven effects.

%
%xxxxxxxxxxxxxxxxxxxxxxxxxxxxxxxxxxxxxxxxxxxxxxxxxxxxxxxxxxxxxxxxxxxxxxxxxxxxxx
\begin{acknowledgments}
We thank F.H.L. Essler and H. Katsura for correspondence.
Financial support from the Laurentian University Research
Fund (LURF) is gratefully acknowledged.
\end{acknowledgments}
%xxxxxxxxxxxxxxxxxxxxxxxxxxxxxxxxxxxxxxxxxxxxxxxxxxxxxxxxxxxxxxxxxxxxxxxxxxxxxx
%

%~\\

%%%%%%%%%%%%%%%%%%%%%%%%%%%%%%%%%%%%%%%%%%%%%%%%%%%%%%%%%%%%%%%%%%%%%%%%%%%%%%
%%%%%%%%%%%%%%%%%%%%%%%%%%%%%%%%%%%%%%%%%%%%%%%%%%%%%%%%%%%%%%%%%%%%%%%%%%%%%%
\bibliography{CondMattRefs}
%\bibliographystyle{apsrev4-1} \bibliography{ref2}
%xxxxxxxxxxxxxxxxxxxxxxxxxxxxxxxxxxxxxxxxxxxxxxxxxxxxxxxxxxxxxxxxxxxxxxxxxxxxxx
%
\end{document}